\documentclass{jfm}
\usepackage{graphicx}
\usepackage{color}
\usepackage{epstopdf, epsfig}
\usepackage{psfrag}
\usepackage[intlimits]{amsmath}

\usepackage{amssymb}
\usepackage{amsmath}
\usepackage{upgreek}
\usepackage{enumitem}
\usepackage[toc,page]{appendix}
\usepackage{float}
\usepackage{tabu}
\usepackage{array}
\usepackage{subcaption}
\usepackage[colorlinks=true, linkcolor=blue, citecolor = blue]{hyperref}
\usepackage{xcolor}
\newcommand{\myRed}[1]{\textcolor{black}{#1}}

\usepackage{blindtext}
\shorttitle{Numerical Simulations of low-speed viscoelastic jets}
\shortauthor{K. Zinelis et al.}

\title{Transition to elasto-capillary thinning dynamics in viscoelastic jets}

\author{Konstantinos Zinelis \aff{1,2}, Thomas Abadie \aff{1,3}   Gareth H. McKinley \aff{2}
  \and Omar K. Matar\aff{1}\corresp{\email{o.matar@imperial.ac.uk}}}
  
\affiliation{
\aff{1}Department of Chemical Engineering, Imperial College London, London SW7 2AZ, UK
\aff{2}Department of Mechanical Engineering, Massachusetts Institute of Technology, Cambridge, MA 02139, USA
\aff{3}School of Chemical Engineering, University of Birmingham, Birmingham B15 2TT, UK
}

\begin{document}

\maketitle
\begin{abstract}
We perform simulations of an impulsively-started, axisymmetric viscoelastic jet exiting a nozzle and entering a stagnant gas phase using the open-source code {\it Basilisk}. This code allows for efficient computations through an adaptively-refined volume-of-fluid technique that can accurately capture the deformation of the liquid-gas interface. We use the FENE-P constitutive equation to describe the viscoelasticity of the liquid and employ the log-conformation transformation, which provides stable solutions for the evolution of the conformation tensor as the jet thins down under the action of interfacial tension. For the first time, the entire jetting and breakup process of a viscoelastic fluid is simulated, including the pre-shearing flow through the nozzle, which results in an inhomogeneous initial radial stress distribution in the fluid thread that affects the subsequent breakup dynamics. The evolution of the velocity field and the elastic stresses in the nozzle are validated against analytical solutions where possible, and the early-stage dynamics of the jet evolution are compared favourably to the predictions of linear stability theory. We study the effect of the flow inside the nozzle on the thinning dynamics of the viscoelastic jet (which develops distinctive ``beads-on-a-string" structures) and on the spatio-temporal evolution of the polymeric stresses in order to systematically explore the dependence of the filament thinning and breakup characteristics on the initial axial momentum of the jet and the extensibility of the dissolved polymer chains.
\end{abstract}
\section{Introduction\label{Intro}}
Spray formation, which involves the disintegration of  a continuous liquid stream as it enters into a stagnant gaseous phase, is an important aspect of many industrial and biological processes \citep{Villermaux2007}. Some representative examples include inkjet printing processes \citep{Basaran2013, Lohse2022}, the dispersal of fertilizers and pesticides on plants \citep{Xu2021}, as well as human sneezing \citep{Scharfman2016}. In particular, sprays are the result of an atomisation process in which a liquid jet is destabilized and undergoes breakup into ligaments, threads, and eventually droplets, under the action of capillary, inertial, and viscous forces. The breakup process involves complex interfacial topological transitions featuring pinch-off singularities where the filament radius locally goes to zero. 

The addition of polymers to a Newtonian solvent endows the resultant polymeric solution with elasticity, which can significantly influence these destabilization and breakup processes; due to the increased extensional resistance to elongation that arises as a result of stretching of the polymeric chains. Examples of viscoelastic systems include paints, inks, industrial thickeners, anti-misting polymer agents, and human saliva or mucus. Achieving a fundamental understanding of the fragmentation process in the presence of viscoelastic effects will facilitate optimisation and control of the droplet size distribution associated with sprays of polymeric solutions. 
    
In this paper, we first consider the phenomenology of the  thinning and breakup of thin filaments of a polymeric solution that leads to beads-on-a-string (BOAS) structures, for which there is no analogue in simple fluids; these structures correspond to a series of almost cylindrical filaments connecting spherical beads \citep{Clasen2006}. Extensive  experimental and numerical work has been conducted to understand the processes leading to the formation of BOAS structures in a thinning viscoelastic filament. The typical formulation makes use of slender jet profile approximation, resulting in a one-dimensional description \citep{Clasen2006, Eggers2008} of the jet radius, axial velocity, and polymeric stress components. More recently, the self-similar profiles of a viscoelastic thread during the thinning process have also been computed with Direct Numerical Simulations (DNS) \citep{Turkoz2018, Snoeijer2020}, highlighting the local importance of the polymeric shear stress components to the final pinch-off.
    
A characteristic exponential rate-of-thinning in the viscoelastic filament, in which the initially capillary-driven deformation of the fluid interface is eventually balanced by fluid elasticity, has also been observed both experimentally \citep{Bazileveskii1990, Entov1997, Amarouchene2001, Clasen2006, Deblais2020} and through numerical simulations \citep{Bousfield1986, Etienne2006, Bhat2008, Bhat2010a, Morrison2010, Turkoz2018, eggers_herrada_snoeijer_2020}. The analytic studies have primarily considered infinitely-extensible polymeric chains that can be described by the Oldroyd-B \citep{Bird1987} constitutive equation  \citep{Li2003, Clasen2006, Ardekani2010}, while computational simulations have  investigated the final breakup of the thread that results when finite extensibility of macromolecules is incorporated by using the FENE-P \citep{Bird1987} model to describe the polymeric stress evolution \citep{Anna2001, Fontelos2004, Wagner2005}. In particular, the role of finite extensibility in controlling the final breakup time (or length) of viscoelastic threads has been examined, and an  analytical solution, which describes the local pinch-off dynamics, has been derived \citep{Entov1997, McKinley2005, Wagner2015}. In addition, the generation of satellite droplets, as well as the influence of viscosity and the fluid relaxation time on a viscoelastic filament initially at rest have been studied via two-dimensional simulations \citep{Liu2023}. Recently, numerical simulations have also been performed of viscoelastic jets and single droplet breakup in an inkjet-printing configuration (without considering the flow in a nozzle) using the Oldroyd-B model \citep{Turkoz2021}, and an adaptive remeshing algorithm was introduced to resolve the time-dependent evolution of very thin viscoelastic threads.
    
To understand the dynamical behaviour and the consequences of the viscoelastic nature of non-Newtonian liquids, we must  quantify the extensional rheological properties of this class of materials. To this end, various experimental protocols accounting for different initial liquid configurations have been developed. The Capillary Breakup Extensional Rheometer (CaBER) \citep{Bazileveskii1990, Yesilata2006} for various dilute and semi-dilute polymeric solutions, and the Rayleigh Ohnesorge Jetting Extensional Rheometer (ROJER) \citep{Keshavarz2015} are the most frequently applied methods to measure the extensional rheology of viscoelastic fluids. The latter  technique is equivalent to a ``flying" CaBER, and allows the examination of a wider range of fluid relaxation times. Additional results for both industrial and biological fluids have also been reported using other recently developed instruments that exploit capillary-driven breakup such as Dripping-onto-Substrate (DoS) Rheometry \citep{Dinic2015, Dinic2019, MartinezNarvaez2021, Lauser2021} as well as for more standard dripping experiments which focus on the transition to elasticity-dominated thinning \citep{Amarouchene2001, Wagner2005, Rajesh2022}. 
    
Although the establishment of an elasto-capillary (EC) balance that results in an exponential rate-of-thinning in the fluid thread has been validated in each of these extensional rheometry configurations, recent experimental and numerical considerations have argued that there can be systematic differences in the local rate of thinning observed in CaBER and ROJER experiments \citep{Mathues2018}. This may be due to non-zero initial values of the axial stresses in the filament which develop as the liquid jet is expelled through the nozzle exit in the ROJER configuration, subtly altering the tensile force balance on a thin viscoelastic filament and leading to a $33 \%$ faster exponential decrease of the local jet radius. These faster dynamics were reported for the first time in the limit of very low jet ejection flowrates due to the so-called ``gobbling phenomenon", with the conventional elasto-capillary balance expected in the CaBER instrument being established for larger flowrate values \citep{Keshavarz2015, Sharma2015a}. The origins of these different thinning dynamics and their dependence on the viscoelastic properties of the fluid and process parameters such as the nozzle radius and the jet velocity remain poorly understood. 
    
The present work aims to address the issues highlighted above and facilitate the design of robust extensional rheometric instrumentation for accurately measuring the rheological properties of weakly elastic complex fluids, such as the relaxation time, the transient extensional viscosity, and the corresponding strain rate. This requires developing a quantitative understanding of the delicate interplay among viscous, inertial, capillary, and elastic forces in low-speed axisymmetric jets in order to establish robust foundations for future studies of the complex dynamics of viscoelastic sprays and the ensuing droplet size distributions that develop at higher flowrates. To achieve this, axisymmetric numerical simulations are carried out over a wide range of system parameters using the open-source code {\it Basilisk} \citep{Popinet2009} which incorporate the viscoelastic shearing flow upstream of the nozzle exit in addition to the subsequent capillarity-driven evolution of the jet and the formation of beads-on-a-string morphologies. \myRed{The azimuthal component of the extra polymeric stress tensor is also accounted for in the current numerical framework, following the formulation by \citet{Lopez-Herrera2019}.}

The rest of this article is organised as follows. In Section \ref{sec:Formulation}, the problem formulation and numerical procedure used to carry out the computations are outlined. A discussion of the results is provided in Section \ref{sec:results} highlighting the stress profiles that develop due to the flow within the nozzle and the subsequent evolution of the jet towards breakup following its exit from the nozzle. Careful attention is paid to the evolution of the jet thinning characteristics with changes in the flowrate and the finite extensibility of the dissolved polymer. Finally, concluding remarks are provided in Section \ref{sec:Conclusions}. 



\section{Formulation and Methodology\label{sec:Formulation}}
We first provide details of the problem formulation which encompasses the flow configurations, the governing equations including the constitutive relation used to describe the fluid viscoelasticity, and the numerical methodology deployed to carry out the simulations.

\subsection{Governing equations and numerical method \label{sec:Equations}}
\begin{figure}
        \centering{\includegraphics[width=1\textwidth]{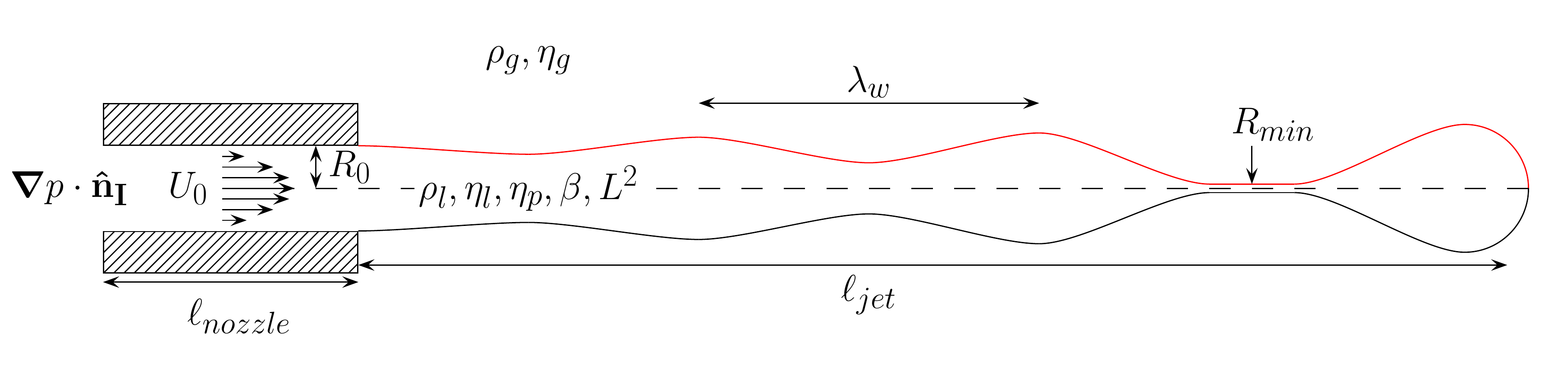}}
        \caption{Schematic representation of the flow depicting the jetting and eventual breakup of a viscoelastic fluid issuing from a nozzle surrounded by a gaseous phase. The flow upstream of the nozzle exit is fully-developed, driven by an applied pressure gradient. The solution domain is highlighted in red.} 
        \label{fig:sketch_jet}
    \end{figure}

The simulation setup for an axisymmetric jet of an incompressible fluid of density $\rho_l$ issuing from a nozzle of length $\ell_{nozzle}$ and initial radius $R_{0}$, is presented in Figure \ref{fig:sketch_jet}. The fluid corresponds to a viscoelastic polymer solution whose polymeric chains have finite extensibility $L$; the total fluid zero-shear viscosity is $\eta_0=\eta_p+\eta_s$ wherein $\eta_p$ and $\eta_s$ denote the polymer and solvent contributions to the viscosity, respectively. Additionally, $\beta$ is the fluid viscosity ratio $\eta_s / \eta_0$ which determines the relative polymeric and solvent contributions to the total dynamic viscosity of the polymer solution. The jet is surrounded by a gas of density $\rho_g$ and viscosity $\eta_g$, and gravitational effects are neglected in the present work. In the simulations, we take the characteristic liquid viscosity to be $\eta_l \equiv\eta_0$, where $\eta_l$ stands for the viscosity of the liquid phase. A pressure gradient (see section \ref{sec:Numerical} below) is imposed along the nozzle in the axial direction $x$ (here, the considered axisymmetric coordinates are $(r, \theta, x$)) leading to the development of a parabolic velocity profile with mean velocity $U_{0}$ within the nozzle. 
As shown schematically in Figure \ref{fig:sketch_jet}, following its exit from the nozzle, the jet evolves downstream with a characteristic wavelength $\lambda_{w}$ due to the imposed perturbations, undergoing an increasingly pronounced deformation over a length $\ell_{jet}$ until ultimately a breakup event occurs. Jet breakup proceeds via the development of a thin thread that connects the leading drop with the rest of the liquid core and polymer stresses in this highly stretched thread act to retard its eventual detachment. Following a fluid element of fixed Lagrangian identity in this thinning ligament reveals an exponential decrease in the radius with time. As will be discussed below, this exponential local decrease of $R_{min}(t)$ results from the establishment of a local elasto-capillary balance in the thinning thread. 

The jet dynamics are governed by the one-fluid formulation of the continuity and momentum equations, which are respectively expressed by:
%
    \begin{equation}
    \boldsymbol{\nabla \cdot \mathbf{u}} = 0, 
    \label{eq:continuity}
    \end{equation}
    \begin{equation}
    \rho\left(\frac{\partial \mathbf{u}}{\partial t}+\mathbf{u} \cdot \boldsymbol{\nabla} \mathbf{u}\right)=-\boldsymbol{\nabla}  p+\boldsymbol{\nabla} \cdot \boldsymbol{\sigma}+\gamma \kappa \mathbf{n} \delta_{s}, 
    \label{eq:NS}
    \end{equation}
where $t$, $\rho$, $\mathbf{u}$, $p$, $\boldsymbol{\sigma}$, $\gamma$, $\kappa$, $\mathbf{n}$, and $\delta_{s}$ stand for time, local density, velocity, pressure, the total stress, (constant) surface tension, interfacial curvature, the outward-pointing unit vector to the interface, and the Dirac delta function (zero everywhere except at the interface), respectively. Here, $\gamma\kappa{\bf n}\delta_s$ denotes the surface tension forces distributed in the cells in the vicinity of the interface with the Continuum Surface Force method \citep{Popinet2009, Popinet2018}. 
Given the viscoelastic nature of the fluid, the total stress is defined as the sum of the solvent and polymeric stresses,  $\boldsymbol{\sigma}=\boldsymbol{\sigma}_{s} + \boldsymbol{\sigma}_{p}$, where $\boldsymbol{\sigma}_{s}=\eta_{s}(\boldsymbol{\nabla} \mathbf{u}+( \boldsymbol{\nabla}\mathbf{u})^{T})$ is the viscous contribution to the total stress tensor, and $\boldsymbol{\sigma}_{p}$ is the polymeric stress tensor defined in the present work by the FENE-P constitutive equation:
        %
        \begin{equation}
        \boldsymbol{\sigma}_{p}=\frac{\eta_{p}} {\tau} \left( \frac{\mathbf{A}}{1-\frac{\operatorname{tr}(\mathbf{A})} {L^2}} 
        -\mathbf{I} \right);
        \label{eq:stress}
        \end{equation}
         %
here $\tau$ is the single characteristic relaxation time of the viscoelastic fluid, $L^2$ provides a measure of the finite extensibility of the polymeric chains, and $\mathbf{A}$ is the dimensionless conformation tensor of the finitely extensible nonlinearly elastic (FENE) dumbbells that model the evolution of the polymer configuration, which is governed by the following equation:
    %
    \begin{equation}
    \frac{\partial \mathbf{A}} {\partial t} + \mathbf{u} \cdot \boldsymbol{\nabla} \mathbf{A} -\left(\boldsymbol{\nabla} \mathbf{u} \cdot \mathbf{A}+ \mathbf{A} \cdot \boldsymbol{\nabla} \mathbf{u}^{T}\right)= -\frac{1}{\tau} \left(\frac{\mathbf{A}}{1-\frac{\operatorname{tr}(\mathbf{A})}{L^2}}
    -\mathbf{I} \right).
    \label{eq:conformation}
    \end{equation}
     %

The above equations are rendered dimensionless by using  the nozzle radius $R_0$, the Rayleigh velocity $U_R= \sqrt{\gamma / \left( \rho_l R_0 \right)}$, the Rayleigh time scale $t_R = R_0/U_R = \sqrt{\rho_l R_0^3/\gamma}$, and the capillary pressure $\rho_{l} U_R^2 = \gamma/R_0$ as the characteristic length, velocity, time, and pressure/stress scales, respectively. Introduction of this scaling into Eqs. (\ref{eq:continuity})-(\ref{eq:conformation}) leads to the following dimensionless equations:
\begin{equation}
    \boldsymbol{\tilde{\boldsymbol{\nabla}} \cdot \mathbf{\tilde{u}}} = 0,
    \label{eq:scaled-continuity}
\end{equation}
\begin{equation}
     \Tilde{\rho} \left(\dfrac{\partial \mathbf{\Tilde{u}}}{\partial \Tilde{t}} +  \mathbf{\Tilde{u}} \cdot \Tilde{\boldsymbol{\nabla}}\mathbf{\Tilde{u}} \right)
  = - \Tilde{\boldsymbol{\nabla}} \Tilde{p} + Oh \left(\beta \Tilde{\boldsymbol{\nabla}}\cdot \boldsymbol{\Tilde{\sigma}}_s + \frac{\left(1-\beta \right)}{De} \Tilde{\boldsymbol{\nabla}} \cdot \boldsymbol{\Tilde{\sigma}}_p \right)
  +\Tilde{\kappa} \Tilde{\delta} \mathbf{n} ,
    \label{eq:scaled_NS}
\end{equation}
\begin{equation}
    \frac{\partial \mathbf{A}} {\partial \Tilde{t}} + \mathbf{\Tilde{u}} \cdot \boldsymbol{\Tilde{\boldsymbol{\nabla}}} \mathbf{A} -\left(\boldsymbol{\Tilde{\boldsymbol{\nabla}}} \mathbf{\Tilde{u}} \cdot \mathbf{A}+ \mathbf{A} \cdot \boldsymbol{\Tilde{\boldsymbol{\nabla}}} \mathbf{\Tilde{u}}^{T}\right)= - \frac{1}{De}
 \left(\frac{\mathbf{A}}{1-\frac{\operatorname{tr}(\mathbf{A})}{L^2}}
    -\mathbf{I} \right), 
    \label{eq:scaled-conformation}  
\end{equation}
where $De=\tau/(R_0/U_R)=\tau \sqrt{\gamma/\rho_l R_0^3}$ denotes the Deborah number, which represents the ratio of the polymer relaxation time to the Rayleigh time scale, and $Oh=\eta_l/\sqrt{\rho_l\gamma R_0}$ is the Ohnesorge number that reflects the competition between capillary, inertial, and viscous forces. The tildes designate dimensionless variables.

Additionally, the importance of the polymer elasticity in the neck can also be understood through a local strain rate  $\dot{\epsilon}_{min} =-2 D \left(\log(R_{min}^{[\alpha]})\right) /Dt$ for each local minimum radius observed in each neck that is formed and subsequently develops into a thin cylindrical ligament between two consecutive primary beads (as depicted in Figure \ref{fig:sketch_jet}), where the primary beads are labelled as $[\alpha] = $ A,B,C,... (starting from the beads furthest from the nozzle). When scaled with $\tau$, this local strain rate in a material element, as it is advected downstream,  corresponds to the local Weissenberg number in the thinning ligament:
\begin{equation}
    Wi =  -2 \tau \frac{D\log(R_{min}^{[\alpha]})}{Dt}.
    \label{eq:Wi}
\end{equation}
This definition will be used to characterise the local rate of jet thinning in each neck that develops along the corrugated jet. When capillarity and elasticity locally govern the dynamics in the elasto-capillary regime as breakup is approached it is expected that the Weissenberg number will approach a constant value $Wi \rightarrow 2/3$ \citep{entov_yarin_1984,Bazileveskii1990, Entov1997,Tirtaatmadja2006}. 

As a result of the coil-stretch transition, the local polymeric stresses in a fluid element typically exhibit a steep increase under the influence of extensional deformations at $Wi \geq 0.5$. In such configurations, numerical challenges can emerge during the computation of the polymeric stress tensor and its divergence which is required for Eq. (\ref{eq:stress}). This is known as the High-Weissenberg Number Problem (HWNP) \citep{Renardy2000}. To circumvent the occurrence of any undesired numerical instability, the log-conformation transformation \citep{Fattal2005} is employed which introduces the logarithmic function $\boldsymbol{\Uppsi}$ of the conformation tensor $\mathbf{A}$, which can be computed via a diagonal transformation such that: 
\begin{equation}
    \mathbf{A} = \mathbf{R}\ \boldsymbol{\Uplambda}\ \mathbf{R}^{T}, 
    \label{eq:A}
\end{equation}
\begin{equation}
    \boldsymbol{\Uppsi} = \log \mathbf{A} = \mathbf{R}\ \log\boldsymbol{\Uplambda}\ \mathbf{R}^T\label{eq:Psi}
\end{equation}
where $\boldsymbol{\Uplambda}$ is the diagonal matrix containing the principal eigenvalues of $\mathbf{A}$. \myRed{The numerical procedure for the solution of the transport equation for the conformation tensor $\mathbf{A}$ when the log-conformation transformation is applied, is provided by \citet{Lopez-Herrera2019}}.

A Volume-of-Fluid (VOF) approach \citep{Popinet2009} is used to capture the deforming jet interface by advecting the volume fraction $c$ of the liquid phase in every computational cell; the advection equation for $c$ is given by 
\begin{equation}
     \frac{\partial c }{\partial \Tilde{t}}+\mathbf{\Tilde{u}} \cdot \boldsymbol{\Tilde{\boldsymbol{\nabla}}} c=0.
     \label{eq:c}
     \end{equation}
Using the VOF formulation, which corresponds to a one-fluid approach to solve the two-phase flow, the density $\rho$ and viscosity $\eta$ are then respectively given by \begin{eqnarray}
    \Tilde{\rho} & = c + (1-c)\dfrac{\rho_{g}}{\rho_l},\\
    \Tilde{\eta} & = c + (1-c) \dfrac{\eta_{g}}{\eta_l};
    \label{eq:properties}
    \end{eqnarray}
here, the characteristic density is chosen to be $\rho_l$, and the tilde decoration is suppressed henceforth for brevity. 
The open-source code \emph{Basilisk} \citep{Popinet2009, Turkoz2018, Lopez-Herrera2019, Turkoz2021} is used here to carry out the computations. A piecewise linear interface calculation (PLIC) technique is used for reconstructing the interface \citep{Popinet2009, Lopez-Herrera2019}. The surface tension-dominated flow disintegration of a liquid jet is modelled with high accuracy thanks to the well-balanced numerical discretisation combined with the height-function method to calculate the geometrical properties of the interface \citep{Popinet2009, Popinet2018}. 

\subsection{Numerical set-up\label{sec:Numerical}}
The simulation domain for the axisymmetric simulations of the jet is a plane of dimensions $100 R_0 \times 100 R_0$ as shown in Fig. \ref{fig:sketch_jet}. The bottom boundary is the axial symmetry axis, while zero-gradient Neumann boundary conditions are imposed at the left and right boundaries for all the velocity and polymeric stress components. myRed{At the top boundary, a no-slip Boundary Condition is also considered at $r=10 R_0$ which is located sufficiently far away from the jet region}. A pressure gradient is imposed at the left boundary to drive the fluid through the nozzle while atmospheric pressure is imposed at the right boundary through a Dirichlet condition. For the initial conditions of the jetting process at $t=0$, we consider the liquid initially at rest in the region $0 \leq r \leq R_0$ and $0 \leq x \leq \ell_{nozzle}$, and the dissolved polymers in the fluid to be initially unstretched in the inlet of the nozzle, with \myRed{$A_{xx}=A_{rr}=A_{\theta \theta}=1$}. 

An adaptive mesh refinement (AMR) technique \citep{Popinet2003} following the quadtree-like structure available in \emph{Basilisk} is used to refine the cells based on the location of the interface and the nozzle, as well as in the regions where large gradients of the axial component of the polymeric stress occur. Specifically, starting from a base grid resolution of $8 \times 8$ square cells for the entire domain and refining up to an initial minimum cell size of $\Delta x_{minimum}=0.09$ around the nozzle region, the adaptive scheme refines up to three maximum levels of refinement, \textit{LVL}, gradually increasing from \textit{LVL}=12 to \textit{LVL}=14, which corresponds to a minimum square cell size of $\Delta x_{minimum}=0.02$ and $\Delta x_{minimum}=0.006$, respectively; this provides sufficient resolution to simulate the dynamics accurately as the pinch-off of the filament is approached. More details of the mesh convergence study are provided in Appendix \ref{sec:Appendix_B}. 

\begin{figure}
    \centering
    \subcaptionbox{}{\includegraphics[width=0.49\textwidth]{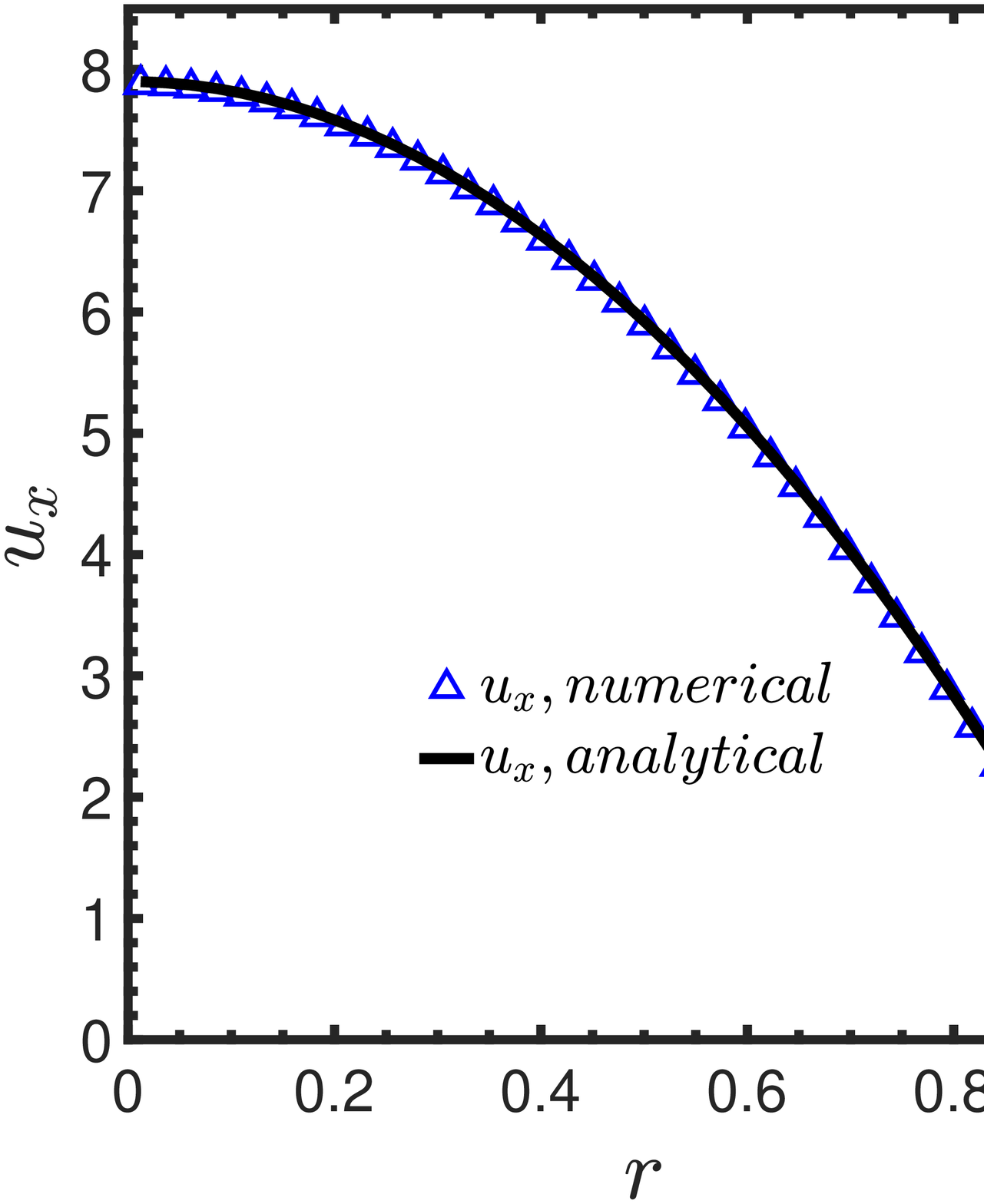}}
    \subcaptionbox{}{\includegraphics[width=0.49\textwidth]{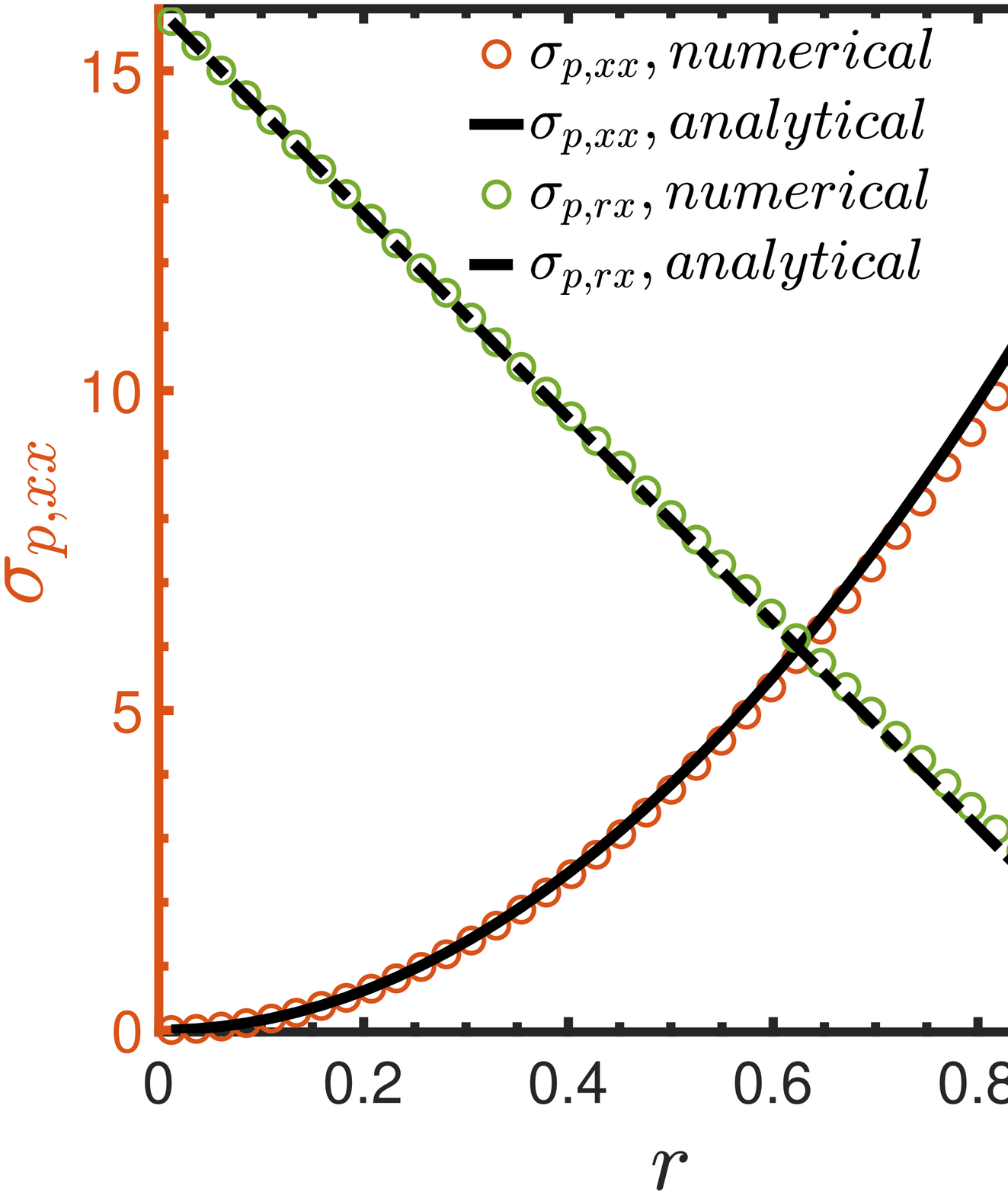}}\\
    \caption{Validation of the predicted fully-developed flow profiles within the nozzle with the parameter values given by Table \ref{tab:params_inside_nozzle}. The analytical solution and corresponding simulation data for the dimensionless axial velocity component and the polymeric stress components are shown in (a) and (b), respectively, as a function of dimensionless position across the nozzle. Here, for the analytical solution \citep{Tome2007, Yapici2009} the dimensionless strain rate is $Wi = \tau U_0 / R_0 = 4$.}
    \label{fig:channel_flow_static}
    \end{figure}

To simulate the flow within the nozzle, a simplified variant of the Immersed Boundary Method (IBM) \citep{Aniszewski2020} is employed to model the velocity field with no-slip and no-penetration conditions imposed at the solid walls. Starting from a static pipe flow case and then continuing with an impulsive injection of the fluid through the nozzle, we first obtain solutions for the velocity and polymeric stress fields and as the flow approaches a steady state we compare the radial profiles of the dimensionless axial velocity field, and axial and shear polymeric stresses as they evolve along the nozzle to the analytical solutions for these fields, which have been derived assuming steady-state and fully-developed axisymmetric flow \citep{Tome2007, Yapici2009}. As we show in Figure \ref{fig:channel_flow_static} (a, b); we obtain excellent agreement between the computed velocity and stress fields and the analytic results.
    \begin{table}
    \caption{Parameter values utilised to validate the flow predictions within the nozzle using the immersed boundary method. 
    }
    \centering
    \begin{tabular}{c c c c c c c }
     \hline
    $De$~~~ & $Oh$ ~~~& $\beta$ ~~~& $L^2$ ~~~& $We$ ~~~& $\ell_{nozzle}$ ~~~& $LVL$\\  [0.5ex]
    1 ~~~& 0.2 ~~~& 0.85 ~~~& $\infty$ ~~~& 16 ~~~& 4 ~~~& 14\\
    \end{tabular}
    \label{tab:params_inside_nozzle}
    \end{table}
    %

Following the validation of the steady pipe flow case, we now investigate the flow inside the nozzle when an axial pressure oscillation with an amplitude $\epsilon_{p}=0.4$ is forced at the inlet using the expression given by Eq. (\ref{eq:perturbation}) with a dimensionless wavenumber $k=0.6$ (with tildes suppressed):
    \begin{equation}
    -{\boldsymbol{\nabla}} {p} \cdot \mathbf{\Hat{n}_I} = 8 Ca \left(1+\epsilon_{p} \sin\left(\sqrt{We}\ k\  {t} \right)\right),
    \label{eq:perturbation}
    \end{equation} 
\noindent    
where $\mathbf{\Hat{n}_I}$ is the unit normal vector to the inlet boundary, $Ca = \eta_l U_{0}/\gamma$ is the capillary number, and $We = \rho_l U_{0}^2 R_0/\gamma$ is the Weber number, which is related to the Ohnesorge number defined in Eq. (\ref{eq:scaled_NS}) through the relation $Ca = \sqrt{We} \ Oh$. 
\myRed{The parameter values used in the simulations are provided in Table \ref{tab:params_inside_nozzle}.} 
The value of $k=0.6$ is selected according to the linear stability analysis presented in Section \ref{sec:LSA} in order to correspond to the most unstable growth rate of the imposed perturbation.

\begin{figure}
            \centering
            \begin{subfigure}{.49\textwidth}
            \includegraphics[width=\textwidth]{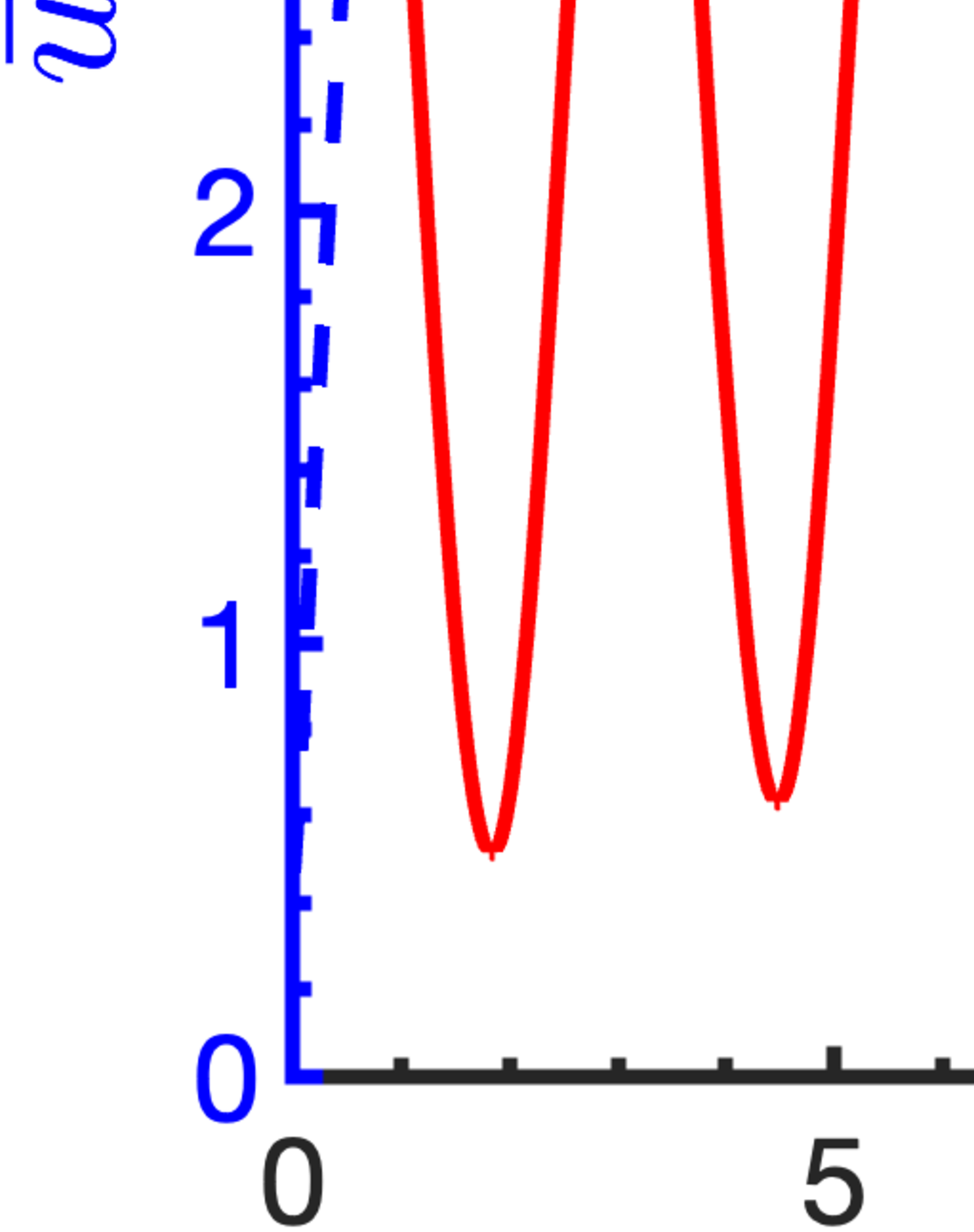}
            \caption{}
            \end{subfigure}
            \begin{subfigure}{.49\textwidth}
            \includegraphics[width=\textwidth]{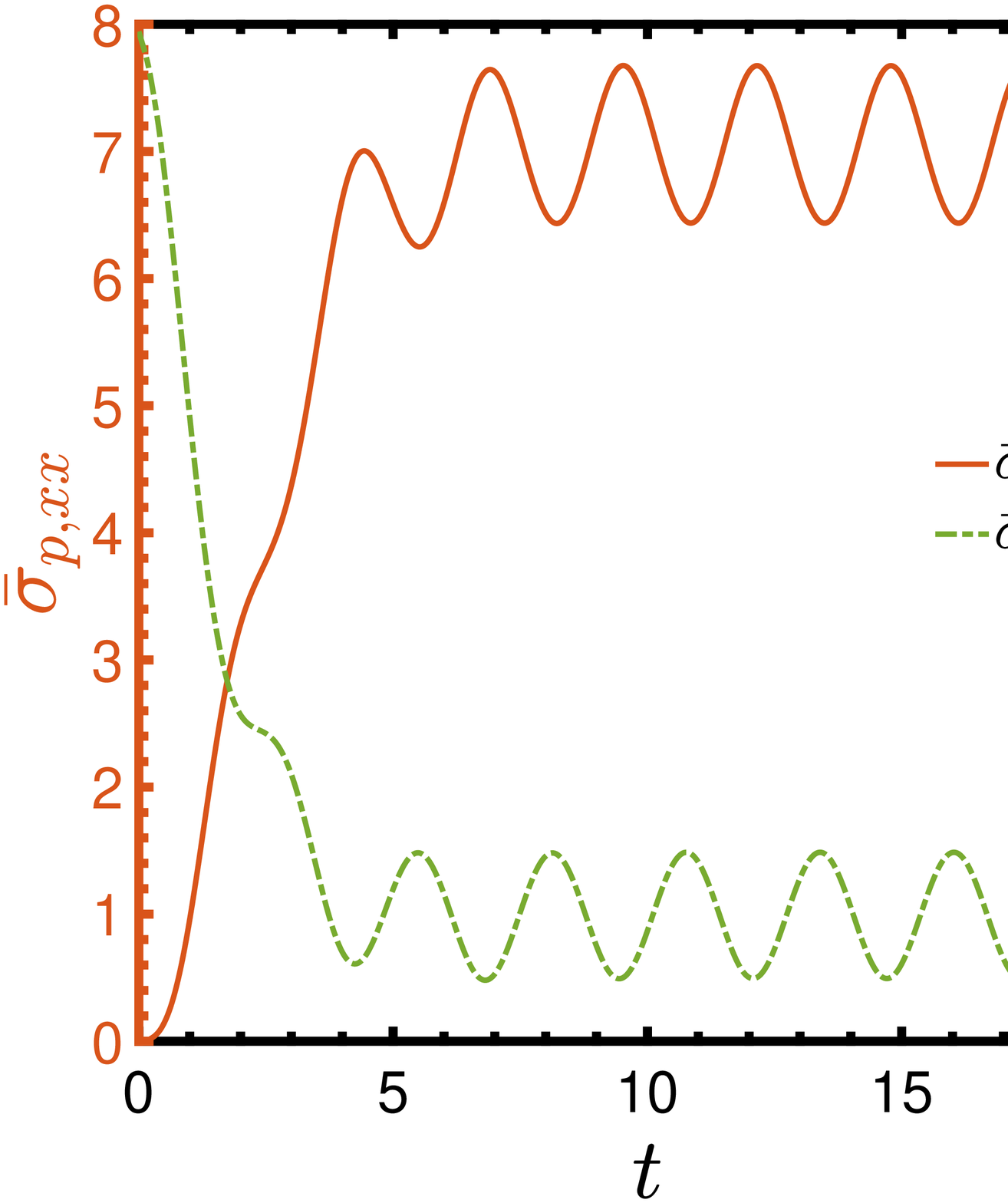}
            \caption{}
            \end{subfigure}
            \caption{Temporal evolution of (a) the cross-sectionally averaged axial velocity component $\bar{u}_x$ %
            and 
            pressure $\bar{p}$;
            and (b) cross-sectionally averaged axial $\bar{\sigma}_{p,xx}$
            and shear $\bar{\sigma}_{p,rx}$
            polymeric stress components in the nozzle. The parameter values are given in Table \ref{tab:params_inside_nozzle} and the pressure gradient is set by Eq. (\ref{eq:perturbation}) with $\epsilon_p=0.4$ and $k=0.6$.}
            \label{fig:channel_flow_pulsed}
        \end{figure}

In Figure \ref{fig:channel_flow_pulsed}(a) we compare the temporal evolution of the radially-averaged magnitudes of the dimensionless isotropic pressure $\bar{p}=\int^1_0 p(r,x=0)rdr$ and axial velocity $\bar{u}_{x}=\int^1_0u_x(r,x=0)rdr$ at a position close to the nozzle inlet. Specifically, Figure \ref{fig:channel_flow_pulsed}(a) indicates the establishment of a phase lag and a corresponding amplitude deviation between the pressure and axial velocity component, as expected from the analysis of \citep{WomerssleyJ.R.1955}. In Figure \ref{fig:channel_flow_pulsed}(b) we also show the evolution of the radially-averaged dimensionless elastic stress components $\bar{\sigma}_{p,xx}=\int^1_0\sigma_{p,xx}(r,x=0)rdr$ and $\bar{\sigma}_{p,rx}=\int^1_0\sigma_{p,rx}(r,x=0)rdr$, validating in both cases the smooth periodic evolution of these quantities. In particular, we show that initially both the polymer shear stress and streamwise axial stress evolve together after the flow is impulsively started at short times $t \leq 5$. However, as time increases, the radially-averaged axial and shear polymeric stress components in the nozzle exhibit distinct trends in magnitude with the dimensionless axial stress significantly increasing and at long times ($t \geq 5$), dominating the shear stress.

        %
%
\begin{table}
\centering
\caption{Simulation parameters for the jetting process. The ratios  $\rho_{g}/\rho_{l}$ and $\eta_{g}$/$\eta_{l}$ are the same as those used by \citet{Turkoz2018}. The dimensionless wavenumber $k$ is selected to be the value that corresponds to the most unstable mode of perturbation expected from linear stability analysis. }
\begin{tabular}{c c c c c c c c c c c c}
\hline
$De$ & $Oh$ & $\beta$ & $L^2$ & $We$ & $\rho_{g}/\rho_{l}$ & $\eta_{g}$/$\eta_{l}$ & $k$ & $\epsilon_{p}$ & $\ell_{nozzle}$ & $\ell_{domain}$ & $LVL$\\  [0.5ex]
1 & 0.2 & 0.85 & 900 & 16 & 0.01 & 0.01 & 0.6 & 0.4 & 4 & 100 & 14\\
\end{tabular}
\label{table_jet}
\end{table}
\begin{figure}
    \begin{center}
    \begin{subfigure}{\textwidth}
        \includegraphics[width=0.85\textwidth]{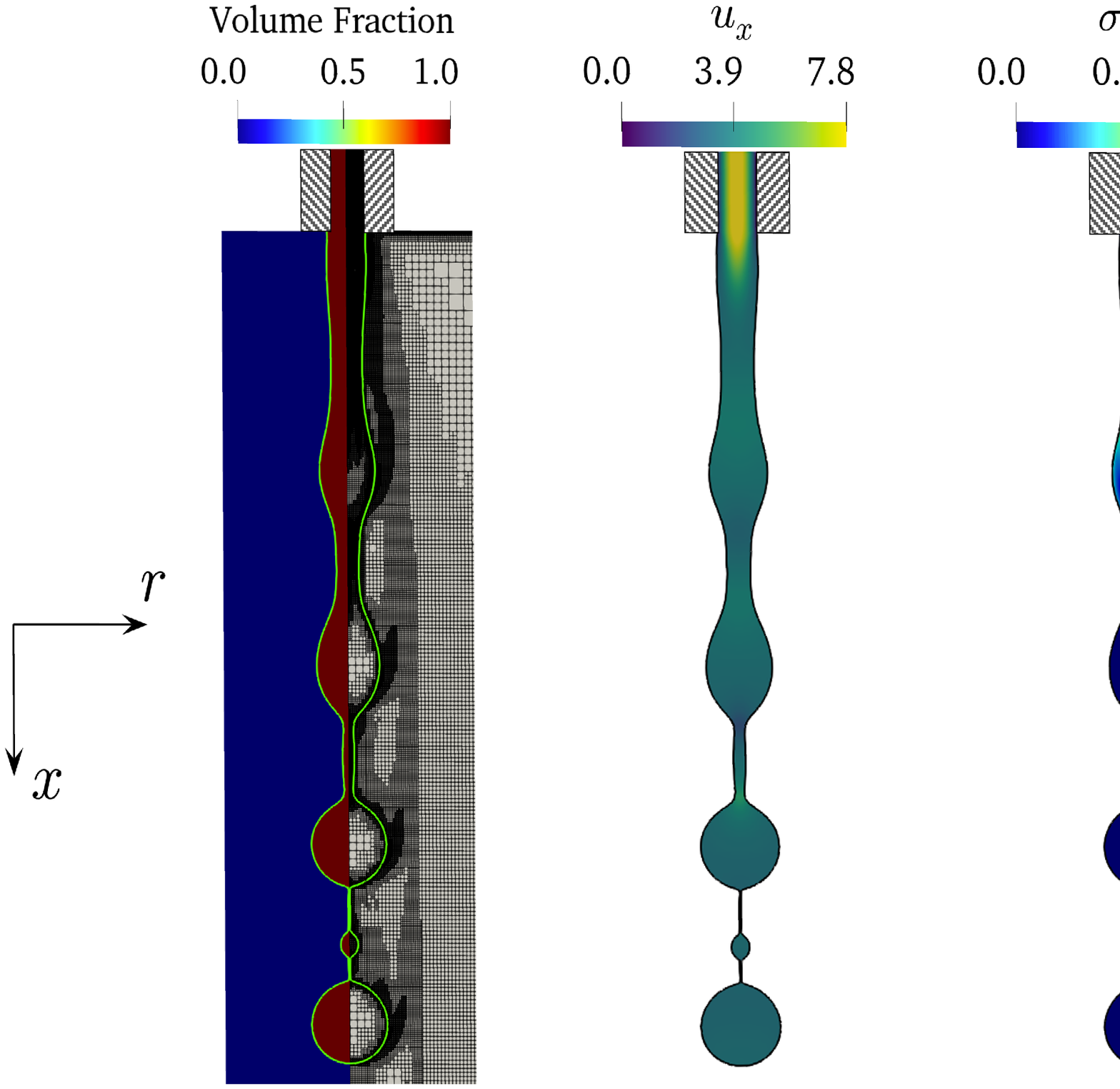}
        \caption{}
    \end{subfigure}
    \begin{subfigure}{0.7\textwidth}
        \includegraphics[width=\textwidth]{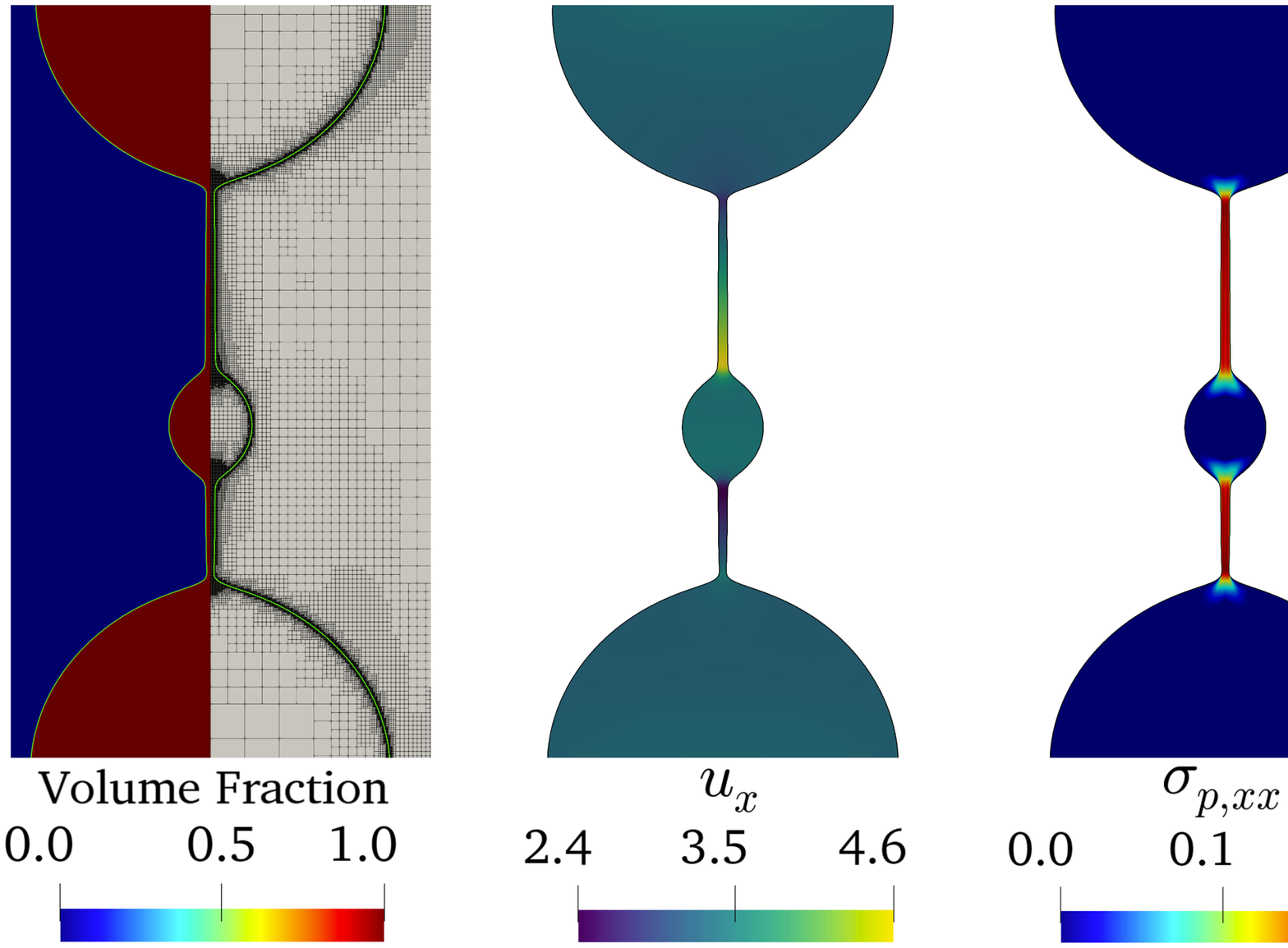}
        \caption{}
    \end{subfigure}
    \end{center}
    \caption{Simulation results at $t=47.8$ for the parameter values listed in Table \ref{table_jet} showing (a) the volume fraction with the mesh highlighting the adaptive mesh refinement in the near-interface regions. The interface is shown with a light green line and the mesh attains its maximum density in the regions which contain thin fluid threads. The spatial distributions of the axial velocity and polymeric stress components are shown in the following contour plots; (b) Local enlargements in the range of $34 \leq x < 41$  of the spatial distributions of the volume fraction with the mesh, the axial velocity, and polymeric stress components in the polymeric thread highlighting the highly localised distribution of the elastic stress and fore-aft asymmetry of the satellite drop that develops far downstream of the nozzle.}
        \label{fig:Simulation_snapshots}
\end{figure}

\section{Results and Discussion\label{sec:results}}

\subsection{Jet evolution and breakup\label{sec:Jetting}}

We present numerical simulations of a low-speed, axisymmetric viscoelastic jet with $De=1$, $Oh=0.2$, $\beta=0.85$, and $We=16$, including the flow within the nozzle, where the mesh resolution is gradually increased in the range $12<LVL<14$ $\left(0.02>\Delta x_{minimum}>0.006\right)$; the rest of the parameters are given in Table \ref{table_jet}. Figure \ref{fig:Simulation_snapshots}(a) shows a contour plot of the volume fraction of fluid in the domain at $t=47.8$, as well as the spatial distribution within the fluid phase of the dimensionless axial components of the velocity, $u_x(r,x)$, and polymeric stress field, $\sigma_{p,xx}(r,x)$. The capillarity-driven deformation of the jet is evident as it exits the nozzle, and this leads ultimately to drop formation. The contour plot of the axial velocity field demonstrates the initial parabolic profile of the axial velocity component which develops upstream of the nozzle exit plane, as expected from Figure \ref{fig:channel_flow_static}(a) discussed in section \ref{sec:Formulation}. This contour plot also shows that material elements at the centerline of the liquid jet leave the nozzle at the maximum velocity and the velocity field rapidly rearranges (within one perturbation wavelength) so it is then advected downstream at a uniform average speed. The spatial distribution of the axial polymeric stresses responds more slowly. Inside the nozzle, the polymeric chains undergo strong shearing close to the wall of the pipe due to the no-slip boundary condition resulting in large stresses. Downstream of the nozzle exit plane, a zero shear stress interfacial condition is imposed, which replaces the no-slip condition on the inside of the nozzle walls. The axial elastic stress component relaxes within the beads that form as the perturbed interface evolves under the action of capillarity, but locally increases in the thin ligaments that develop. This local increase is driven by the large capillary pressure in the filament as it thins towards breakup. 

Figure \ref{fig:Simulation_snapshots}(b) offers a closer inspection of the simulation results in Figure \ref{fig:Simulation_snapshots}(a). In particular, Figure \ref{fig:Simulation_snapshots}(b) reveals that the two leading drops are separated by a thin filament on which a much smaller, satellite drop has formed. These ``beads-on-a-string structures" (BOAS) are the result of a balance between capillarity and the viscoelasticity of the polymer and have no direct analogue in low-speed jets of Newtonian fluids undergoing deformation and breakup. It is also clear that the AMR scheme within \emph{Basilisk} has been deployed appropriately for refinement close to the interface and to resolve accurately the stresses in these thin string-like filaments. We also note that the thread is not perfectly fore-aft symmetric: as time evolves the left upper side of the thread (i.e. the side which is closest to the nozzle) is observed to move slightly faster than the lower side (furthest from the nozzle), and the satellite droplet moves downstream towards the leading drop. Additionally, the dimensionless axial polymeric stresses are seen to attain very high values up to $\text{max}(\sigma_{p,xx}) \approx 50 \gamma / R_0$ in the thin ligaments that develop on either side of the smaller droplet due to the response of the polymer molecules to the elongational flow. Inside the satellite drop and the two bigger drops, the stresses relax to values close to zero.

    \begin{figure}
    \centering
    \begin{subfigure}{0.75\textwidth}
    \centering
    \includegraphics[width=\textwidth]{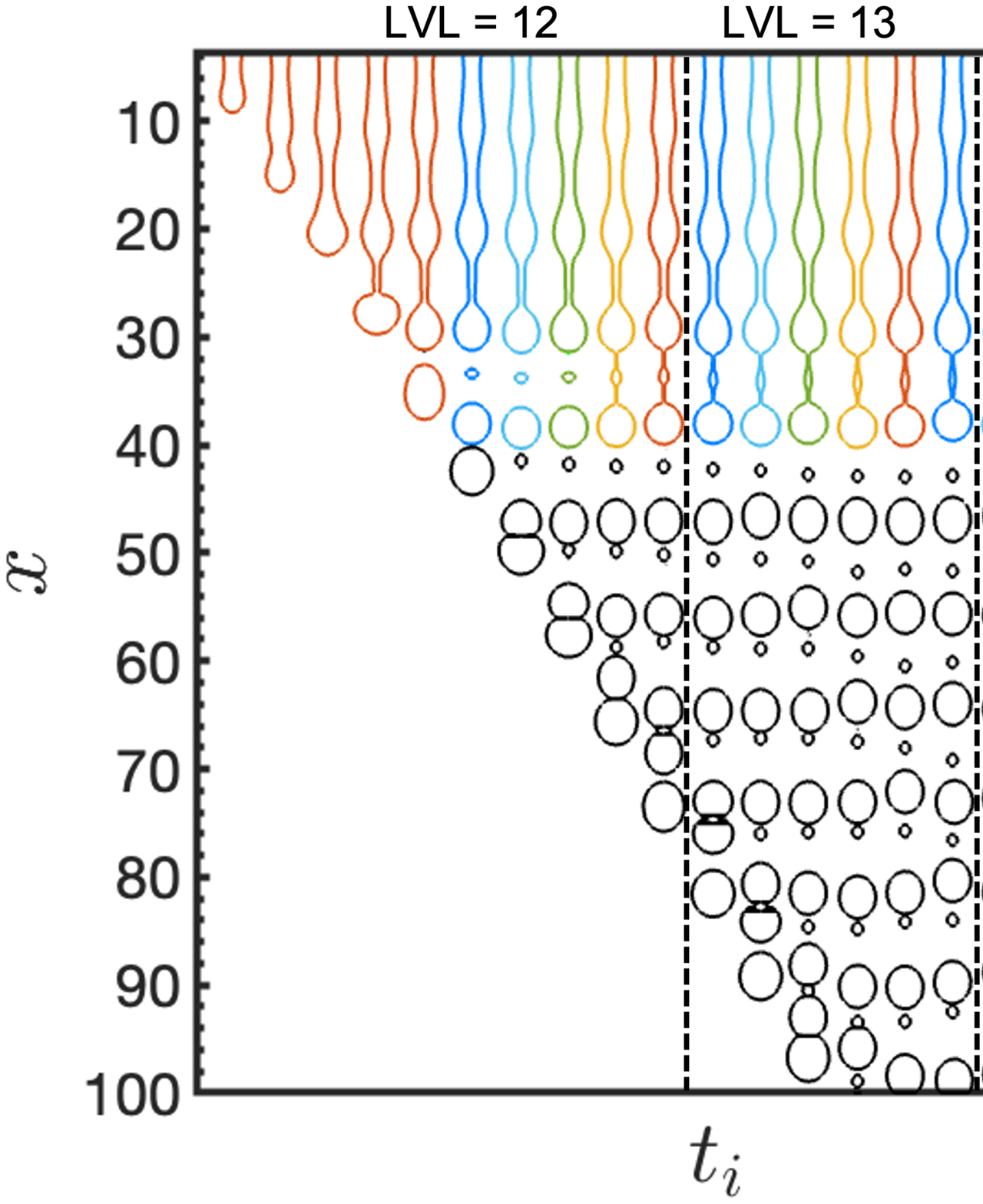} 
    \caption{}
    \label{interface}
    \end{subfigure}
    \begin{subfigure}{0.75\textwidth}
    \centering
    \includegraphics[width=\textwidth]{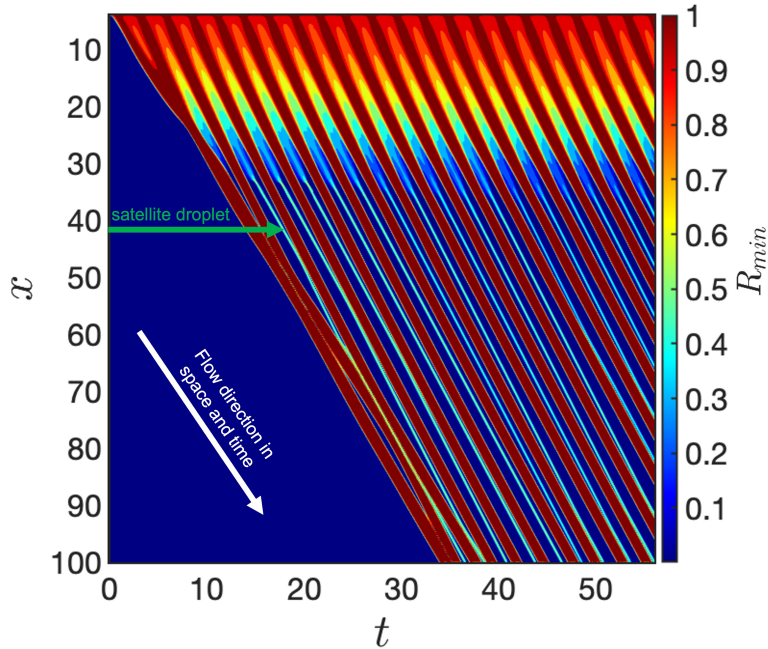}
    \caption{}
    \label{kymograph}
    \end{subfigure}    
        \caption{(a) A sequence of jet profiles $R(x,t_i)$ highlighting the formation of beads-on-a-string structures, as well as primary and satellite drops, captured each period when the velocity attains its minimum values in the inlet (these times denoted $t_i$ can be seen in Figure \ref{fig:channel_flow_pulsed}(a)). (b) Space-time diagram or ``kymograph" showing $R(x,t)$ as each wave pulse is ejected from the nozzle, leading to the formation of a new primary bead as it flows away from the nozzle. The simulation parameters are provided in Table \ref{table_jet}.}
        \label{fig:Interface_kymagraph}
    \end{figure} 
    
Figure \ref{fig:Interface_kymagraph}(a) presents an alternative Lagrangian view of the jetting process. We show the temporal evolution of the interface plotted at a sequence of times denoted $t_i$ when the oscillating velocity forcing of the injection at the inlet attains its minimum value (i.e. $t_i \approx 2.3 + 2 \pi n / (\sqrt{We} \ k) \ \text{with} \ n=0,1,2,...$). Mesh resolution represented by different $LVL$ values increases as time increases to ensure that the dynamics are captured accurately. Each pulsed wave-like disturbance that emanates from the nozzle results in a local necked region (observable by the blue contours in Figure \ref{fig:Interface_kymagraph}(b)) associated with each new emerging primary bead that travels downstream at a constant velocity. After an initial transient period of developing flow, periodicity is observed in terms of the locations where a droplet is formed, as well as where the thin fluid ligament connects the leading droplet to the rest of the jet. Figure \ref{fig:Interface_kymagraph}(b) shows the corresponding ``kymograph'', which highlights the periodicity over the entire spatio-temporal spectrum of the thinning dynamics of the viscoelastic jet. In particular, the contour plot demonstrates how the local minima in the radius of the jet evolve both in space and time, where the colour scale ranges from the initial value of the radius, $R_{0}$, down to the minimum cell size ($ \approx 0.6 \% R_0 $ for $LVL=14$). The kymograph also permits us to track the formation and the detachment of the leading droplet, observed at early times after which periodicity of the jet evolution is established. The spatiotemporal development of satellite droplets, which are represented by cyan-coloured streaks of smaller radius and travel downstream at almost constant speed is also highlighted, while the dark blue regions in the contour plot demonstrate the complete detachment of each of the formed droplets. 
%
\begin{figure}
    \centering
    \includegraphics[width=\textwidth]{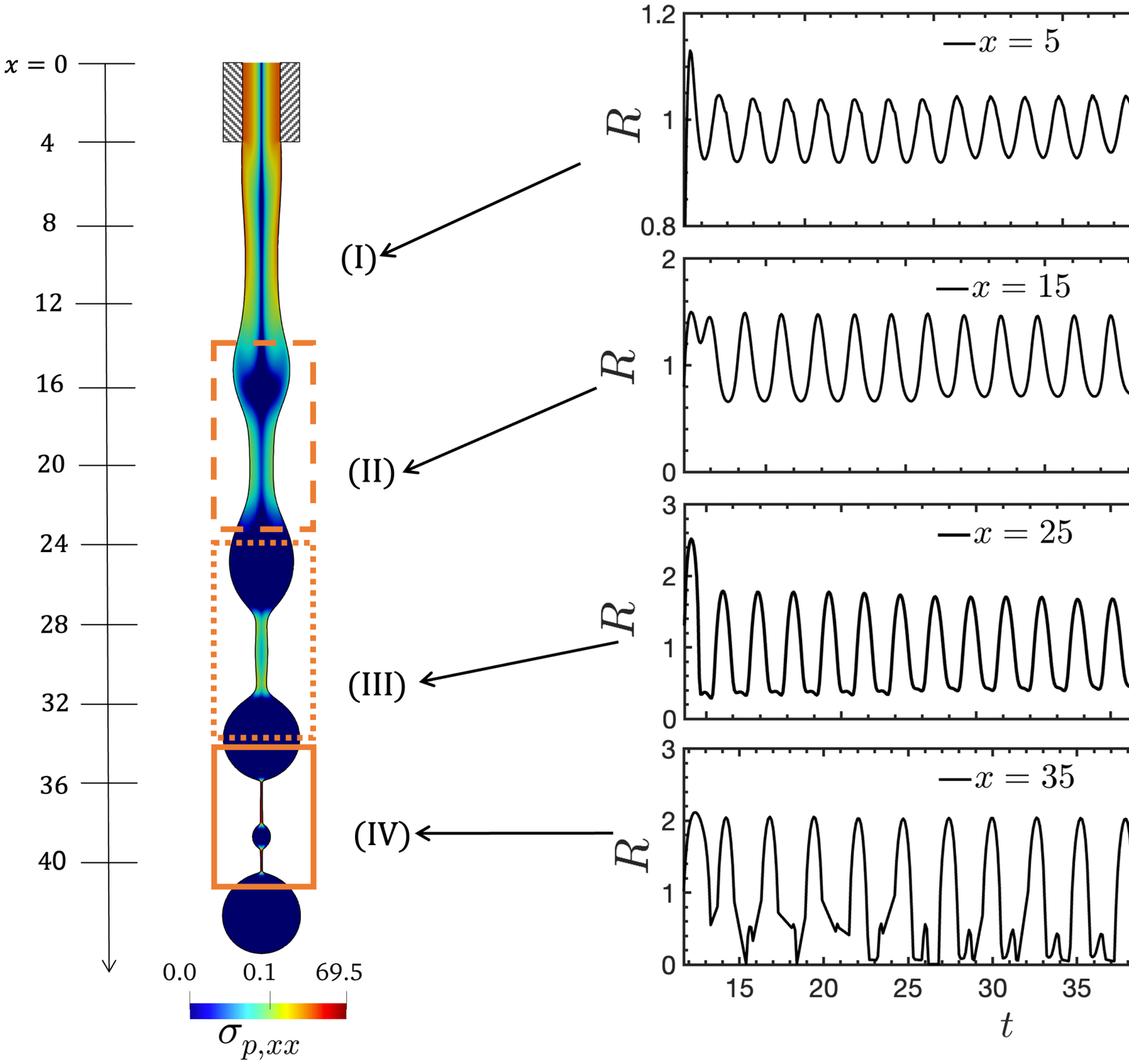}
    \caption{Temporal evolution of the jet radius associated with four distinct regions: `I' and `II' close to the nozzle and at intermediate distances from it wherein the dynamics are linear and weakly nonlinear, respectively. Regions `III' and `IV' are further away from the nozzle where the dynamics are strongly nonlinear featuring the formation of the first satellite beads and beads-on-a-string structures, respectively. The contour plot depicting the shape of the jet at $t=47.8$ in regions `I'-`IV' shows the spatial evolution of the axial component of the polymeric stress tensor accompanied by the corresponding scale of the $x$-axis. The parameter values are the same as in Table \ref{table_jet}.}
    \label{fig:Radius_Signal}
\end{figure}

In Figure \ref{fig:Radius_Signal}, we show the temporal evolution of the jet radius at four fixed Eulerian locations along the jet axis, which correspond to four distinct regions of the breakup process (labelled `I' - `IV') and are characterised by their proximity to the nozzle exit plane and the nature of the evolution of $R(x,t)$. The corresponding jet profile is coloured by the magnitude of the axial component of the polymeric stress tensor. Relatively close to the exit plane of the nozzle (which is located at $x = 4$) in region `I' ($ 4 \leq x < 14$) the jet radius exhibits essentially linear dynamics \citep{Middleman1965, goldin_yerushalmi_pfeffer_shinnar_1969, Brenn2000} characterised by a sinusoidal response to the pressure gradient forcing set by Eq. (\ref{eq:perturbation}). 

Following this linear phase, Figure \ref{fig:Radius_Signal} shows that in a second region denoted  region `II' ($14 \leq x < 24$) the jet radius exhibits a weakly nonlinear behaviour. Further away from the nozzle  ($24 \leq x < 34$)  the radial perturbation strongly enters into a nonlinear regime denoted region `III' featuring bead formation separated by thin filaments, as shown in Figure \ref{fig:Radius_Signal}. The dynamical evolution of $R(x,t)$ at $x=35$ in region `IV' ($34 \leq x < 41$) are strongly nonlinear;  here, in the elasto-capillary regime, the thin ligaments are deformed by capillary forces while elastic stresses delay interface breakup through the development of stable viscoelastic threads and beads-on-a-string structures. It is also clear from a close inspection of Figure \ref{fig:Radius_Signal} that the initial transient response (corresponding to when the first ``leading" droplet exits the nozzle and then passes each Eulerian location) takes longer to decay further from the nozzle but eventually the jet attains a perfectly periodic structure (corresponding to the diagonal lines observed in the space-time diagram of Figure \ref{fig:Interface_kymagraph}).

\begin{figure}
    \centering
    \includegraphics[width=0.85\textwidth]{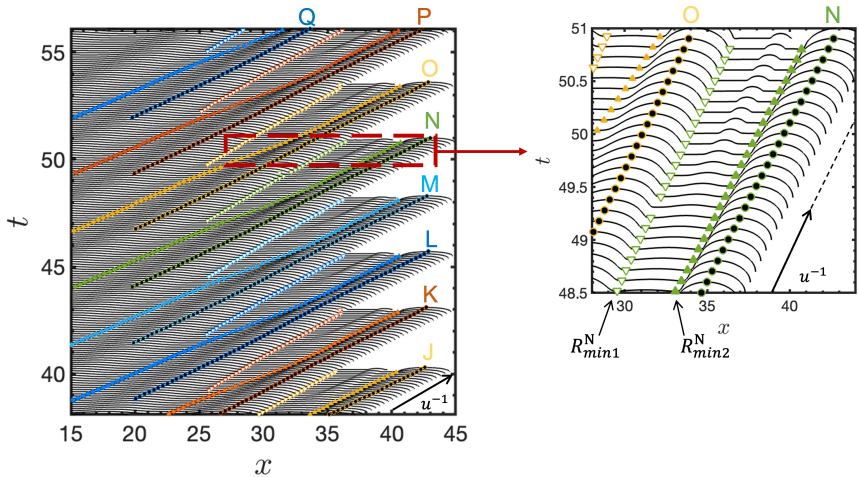}
    \caption{Space-time plot of the spatio-temporal evolution of the jet highlighting the detection and tracking of two different minima of the jet radius on either side of a secondary droplet, always observed for $x \geq 25$. Each local maximum corresponding to the formation of a primary bead is shown by a filled black circle symbol with a distinctly coloured border, as well as a Lagrangian label $[\alpha]$. Each local minimum in the thinning fluid ligament between neighbouring primary beads is shown by a filled (`$R_{min1} ^ {[\alpha]}$ - min1') and hollow (`$R_{min2} ^ {[\alpha]}$ - min2') symbol of the corresponding colour for each neck region established behind each primary bead $[\alpha]$. The enlarged view shows the jet profile between two beads with Lagrangian labels $[\alpha]=$ N, O. The velocity at which the viscoelastic jet evolves downstream is also indicated. The simulation parameter values are the same as in Table \ref{table_jet}.}
    \label{fig:Jet_LSA}
    \end{figure}

In what follows, we first use linear stability theory to study the regions closest to the nozzle, i.e. region `I' and `II', before embarking on a detailed analysis of regions, `III' and `IV', in which the dynamics become increasingly nonlinear. In experimental visualisation of the jet breakup, it is common to follow the evolution of local maxima in the jet radius (leading to the formation of primary beads) as well as local minima in the necks (which ultimately lead to pinch-off). The situation is more complex in viscoelastic jets because the large elastic stresses that develop in the neck can inhibit or totally prevent pinch-off. It is thus important to always follow the same Lagrangian element when attempting to relate local rates of thinning to material properties such as the local extensional viscosity in a material element. To aid our analysis of the nonlinear dynamics, we move to a Lagrangian description so that we follow a specific material element as it is ejected and transported away from the nozzle and is increasingly deformed by capillary effects in space and time. We show in Figure \ref{fig:Jet_LSA} the space-time evolution of the jet interface (oriented horizontally here to conserve space) highlighting the evolution of the wave crests and troughs with distance from the nozzle. This representation allows to ensure that the same minimum is followed in space and time, which is essential for estimating the local instantaneous rate of thinning in region `IV' accurately. As the jet thins and the beads-on-a-string structure fully develops (at $x \simeq 35$), two thin threads are formed, one on each side of the satellite droplet. This results in \textit{two} local minima in the jet radius $R_{min1} ^ {[\alpha]}$ and $R_{min2} ^ {[\alpha]}$ following each labelled primary bead $[\alpha] = $ A,B,C,..., for each period of the upstream forcing. These two emerging minima are henceforth labelled `min1' and `min2' for each local minimum in every neck region.

    %

\begin{figure}
    \centering
    \begin{subfigure}{0.75\textwidth}
    \includegraphics[width=\textwidth]{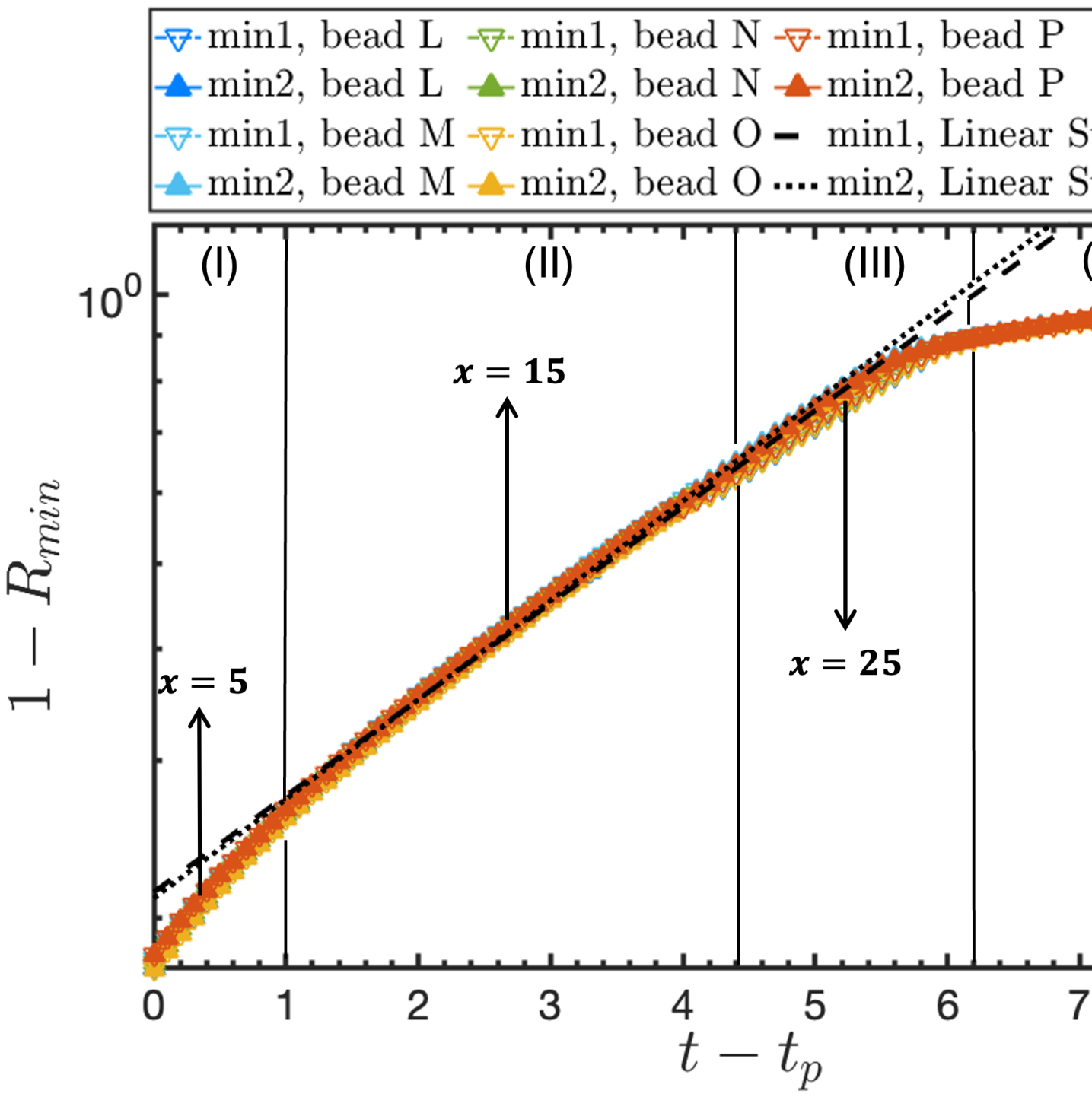}
    \caption{}
    \end{subfigure}
    \begin{subfigure}{0.6\textwidth}
    \includegraphics[width=\textwidth]{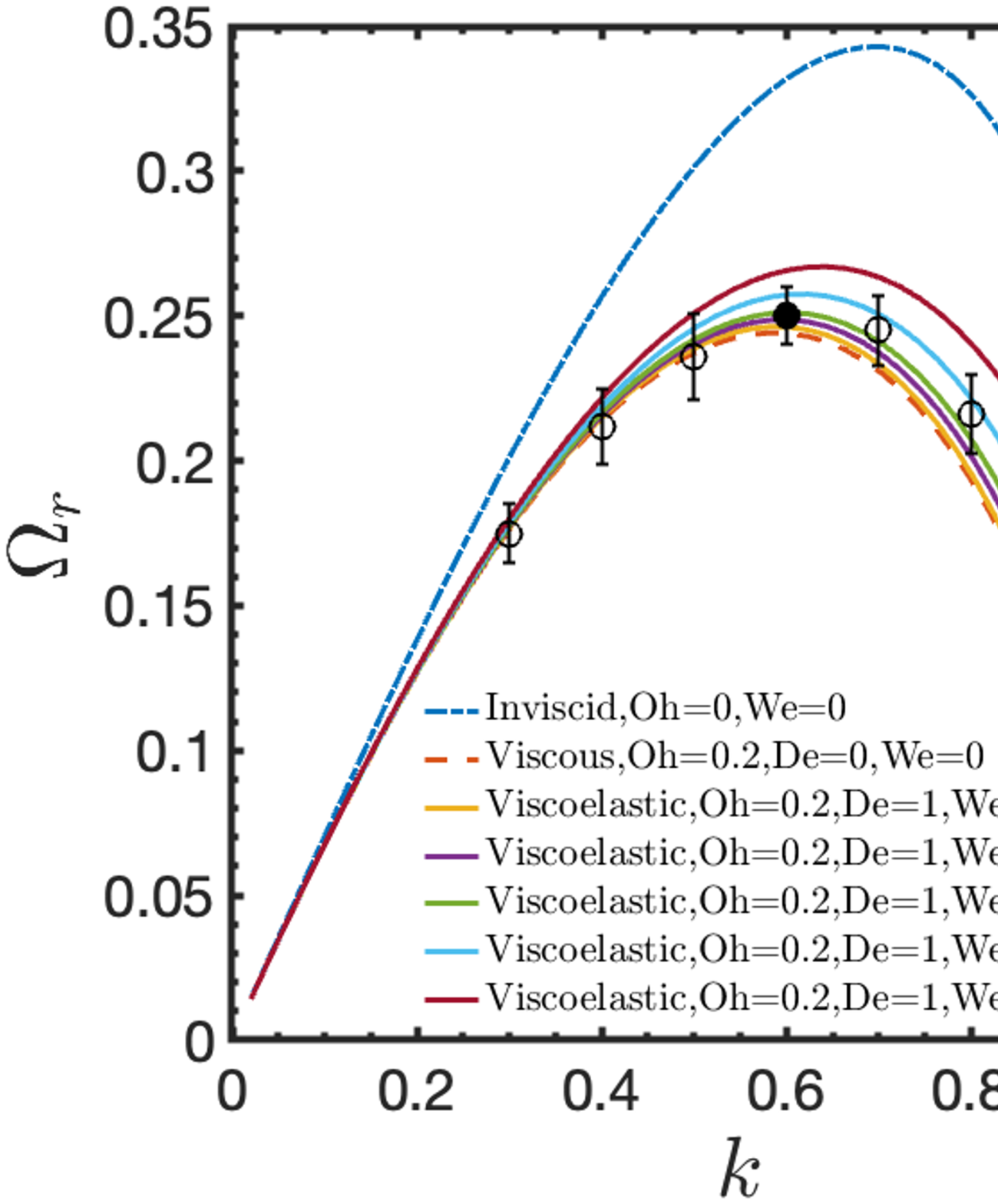}
    \caption{}
    \end{subfigure}
    \caption{Comparison of the numerical predictions with those from a linear stability analysis for regions `I' and `II': (a) semi-log plot of the temporal evolution of the deviation of the jet radius from its base state for an imposed perturbation given by Eq. (\ref{eq:LSA_Eq}) with $\delta=0.13$ and $k=0.6$. The rest of the simulation parameters remain unchanged from Table \ref{table_jet}. Here, $t - t_{p}$ corresponds to the elapsed time since a specific local maximum in the jet radius which coincides with the formation of a primary bead labelled $[\alpha] =$  L, M, N, O, P is formed as a fluid ligament flows away from the nozzle. Regions `III' and `IV' are characterised by nonlinear dynamics for which linear theory is not appropriate; regions `I'-`IV' map onto those identified in Figure \ref{fig:Radius_Signal}. (b) Dispersion curves generated via the solution of Eq. (\ref{eq:LSA_Appendix}) \citep{Middleman1965,Brenn2000} for various  Weber numbers and wavenumbers; simulation data for the specific local neck regions highlighted in Figure \ref{fig:Jet_LSA} were used for the filled circle symbol at $k=0.6$ and the simulation parameters provided in Table \ref{table_jet}; the hollow symbols have been generated for the same $Oh, De$ and $We$ ($Oh=0.2, De=1, We=16$) conditions but different $k$ values.
    }
    \label{fig:nozzle_LSA}
\end{figure}

\subsection{Region `I' and `II': Linear and weakly nonlinear evolution of jet profile \label{sec:LSA}}

In regions `I' and `II', we use linear stability theory to characterise the thinning dynamics. Here, the decrease in the dimensionless jet radius is expected to be described by a growing perturbation of the general form:
\begin{equation}
    R_{min}(t) = 1 - \delta \exp(\Omega t), \myRed{\text{for}}\ t \geq 0
  \label{eq:LSA_Eq}
\end{equation} 
\noindent 
where $\delta$ is the initial perturbation amplitude and $\Omega=\Omega_r + i \Omega_i$ is the complex growth rate which depends parametrically on $Oh$, $De$, and $k$: 
$\Omega_r >0$ indicates the presence of linear instability.
In Figure \ref{fig:nozzle_LSA}(a) we show a semi-log plot of $1-R_{min} ^{[\alpha]}$ as a function of time whence we have subtracted an interval $t_p$ that represents the instant when each of the identified wave pulses, which lead to the formation of primary beads labelled $[\alpha]=$ L, M, N, O, P, exited the nozzle. We also illustrate the locations of the stationary Eulerian points located at 5, 15, 25, and 35 nozzle diameters from the injection inlet. Given the velocity of the jet, each of these fixed points can be associated with a specific value of  $t-t_p$ which corresponds to the time when a specific material element passes through one of each of these locations. Inspection of this plot reveals that even though the perturbations in region `I' are small as illustrated in Figure \ref{fig:Radius_Signal}, the residual stresses in the jet, the rearrangement of the velocity profile in the jet (from parabolic to plug-like) \textit{and} the pinning conditions of the free surface to the nozzle exit at $x=4$ all influence the local growth rate of perturbations. Hence, it is the second region denoted `II' which is best characterised by the linear theory. 

In region `II' (beyond approximately one jet diameter from the nozzle), the exit boundary conditions have been forgotten and small perturbations to the radius grow exponentially under the action of capillary squeezing. Figure (\ref{fig:nozzle_LSA})(a) shows that region `II' remains linear (on a semi-log plot) over a sufficiently large time interval that it is possible to calculate the gradient which can be compared with the dimensionless growth rate $\Omega_r$ in Eq. (\ref{eq:LSA_Eq}). The growth rates obtained from the numerical simulation for $Oh=0.2, De=1$ and $We=16$ and linear stability theory \citep{Brenn2000} (see Appendix \ref{sec:Appendix} for details) are $\Omega_r  \approx 0.246 \pm 0.006$ and 0.25, respectively, for $k=0.6$, demonstrating excellent agreement, as is also shown in Figure \ref{fig:nozzle_LSA}(b).

In particular, \myRed{the polymer finite-extensibility is not expected to exhibit a strong impact at early times of the jet deformation which is capillary-driven and dominated by the linear interfacial disturbances. It only becomes critical as a thread is formed and thins down to considerably smaller lengths \citep{Wagner2015}, as will also be shown in Section \ref{sec:We_effect} We can therefore compare the numerically predicted growth rates with those obtained from linear theory for an Oldroyd-B fluid ($L^2 \rightarrow \infty$) for a range of $k$ values}. Figure \ref{fig:nozzle_LSA}(b) shows the resulting dispersion curves computed using the analysis of \citet{Middleman1965} for a viscoelastic liquid jet in an inviscid gaseous environment. Specifically, the dispersion curve of an inviscid Newtonian jet ($Oh=0, De=0$) with no inertia ($We=0$) \citep{Rayleigh1879} is first presented (blue dashed-dotted line), while viscosity effects are then added for the Newtonian jet ($Oh=0.2, De=0, We=0$) resulting in a strongly reduced growth rate of the perturbation (orange dashed line). Subsequently, the effects of viscoelasticity ($De=1$) are considered (solid lines): first, the influence of inertia is neglected ($Oh=0.2, De=1, We=0$) showing that the linear viscoelastic jet is more unstable (slightly larger maximum growth rate) than a Newtonian fluid of the same viscosity. That is, elasticity contributes to a little faster rate of thinning in the neck of the liquid jet in the linear regime \citep{Middleman1965}.  \citet{Brenn2000} expanded the work of \citet{Middleman1965} by incorporating the effects of the momentum flux arising from the injection of the jet (ie Weber numbers $We>0$. All of the corresponding dispersion curves ($Oh=0.2, De=1, We>0$) exhibit positive growth rates at $k=1$ instead of $\Omega_r=0$ as obtained when $We=0$. In addition, the presence of fluid inertia is destabilising, leading to a higher maximum wave growth rate, a shift of the most unstable mode to larger $k$ values, and a wider range of wavenumbers for which $\Omega_r >0$.

Comparison of the linear theory predictions to the computed growth rates resulting from the numerical simulations for $Oh=0.2$, $De=1$, $We=16$ and $k=0.3,0.4,0.5,0.6,0.7$ and $0.8$ (obtained using a similar procedure to that discussed in connection to Figure \ref{fig:nozzle_LSA}(a)), shows good agreement. Specifically, all the circular symbols in Figure \ref{fig:nozzle_LSA}(b) result from the analysis of the corresponding rates of change, equivalent to the analysis shown in Figure \ref{fig:nozzle_LSA}(a). Analysis of at least five different Lagrangian wave pulses exiting the nozzle result in the error bars shown in the figure. Therefore, the slight deviations seen in Figure \ref{fig:nozzle_LSA}(b) between the simulation predictions and the dispersion curve computed for $Oh=0.2, De=1$ and $We=16$ (green curve) indicate the limits of our ability to resolve small differences in the resulting growth rates due to the small neck-to-neck variations in the profiles of the thinning jet considered in each of the simulations.

   \begin{figure}
    \centering 
    \begin{subfigure}{0.75\textwidth}
        \includegraphics[width=\textwidth]{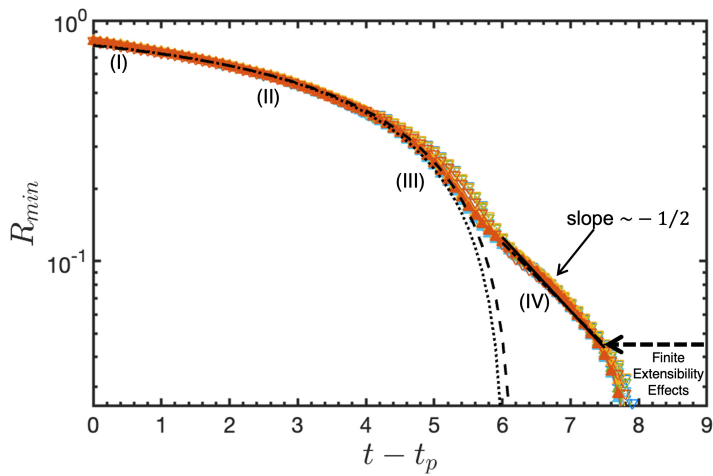}
    \caption{}
    \end{subfigure}
    \begin{subfigure}{0.75\textwidth}
    \includegraphics[width=\textwidth]{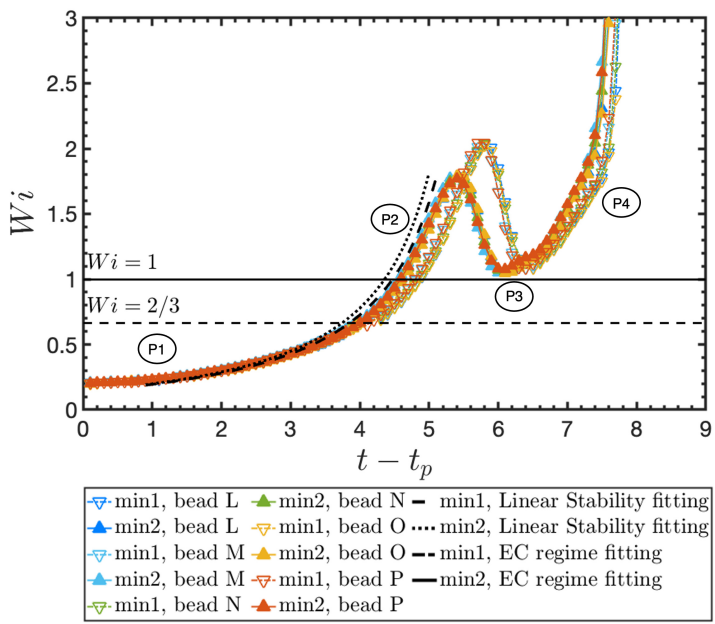}
    \caption{}
    \end{subfigure}
    \caption{(a) Thinning dynamics and (b) the temporal evolution of the local Weissenberg number, $Wi$ in each Lagrangian thinning neck region. Data from each of the local neck regions depicted in Figure \ref{fig:Jet_LSA} are used (for simulation parameters listed in Table \ref{table_jet}). In (b), the plateau values corresponding to $Wi=2/3$ ($\dot{\epsilon}_{min} = 2/ 3 \tau$) and $Wi=1$ ($\dot{\epsilon}_{min} = 1/ \tau$), which coincide with the two distinct limits to the thinning dynamics in the elasto-capillary regime \citep{Keshavarz2015, Mathues2018}, are also shown. The regions `I'-`IV' identified in (a) map directly onto those discussed in Figures \ref{fig:Radius_Signal} and \ref{fig:nozzle_LSA}(a). Points labelled `P1'-`P4' in (b) correspond to the ends of regions `I'-`IV' in (a), respectively.}
    \label{fig:Rmin}
    \end{figure}
    %

\subsection{Regions `III' and `IV': Nonlinear evolution of jet profile \label{sec:EC}}
 
The evolution in jet profile in the nonlinear regimes is first analysed by tracking the temporal decrease of the jet radius following local minima in each neck region between primary beads in a Lagrangian way as it travels downstream away from the nozzle, as shown in Figure \ref{fig:Rmin}(a), for the same simulation parameters as in Table \ref{table_jet}. This highlights the emergence of four distinct regimes, as defined in Figure \ref{fig:Radius_Signal}: `I', `II', `III', and `IV' characterised by $R_{min} \geq 0.75$, $0.75 > R_{min} \geq 0.25$, $0.25 > R_{min} \geq 0.1$ and $0.1 > R_{min} \geq 0.006$, respectively, where $0.6 \% R_{0}$ is the mesh resolution limit according to the maximum $LVL$ value achieved in this work. As the thinning jet enters region `IV' it is clear that the radius of the local minima `min1' and `min2' (corresponding to the local minimum jet radius in each of the two thin ligaments between beads) decreases exponentially in time. This is the elasto-capillary (EC) regime anticipated in the Introduction. Finally, at very small radii below $R_{min} \leq 0.04$, there is a deviation from the exponential thinning corresponding to the onset of finite extensibility effects. In this regime, the thread radius is ultimately expected to thin linearly in time resulting in finite time breakup \citep{Entov1997, Renardy2004}.

To determine the characteristic time-scale that best describes the exponential thinning, the temporal evolution of the local dimensionless strain rate $Wi$, defined in Section \ref{sec:Equations}, is plotted in Figure \ref{fig:Rmin}(b). Specifically,  \myRed{this plot highlights the existence of two different plateau values of $Wi=2/3$ and $Wi=1$ which correspond to the two distinct thinning rates during the elasto-capillary regime as determined by \citet{Keshavarz2015} and \citet{Mathues2018}, respectively.} In addition, Figure \ref{fig:Rmin}(b) is characterised by four points, `P1'-`P4', highlighting the non-monotonic evolution of the local strain rate in the thinning jet in agreement with what has been shown by \citet{Tirtaatmadja2006} and more recently by \citet{Rajesh2022}. The points `P1'-`P4' are respectively associated with the ends of regions `I'-`IV', identified in Figure \ref{fig:Rmin}(a): `P1' is associated with the end of region `I' in which the influence of the exit nozzle on the thinning jet radius is felt, while `P2' coincides with the end of region `II' and exponential growth in disturbances evident in Figure \ref{fig:nozzle_LSA}(a). In the transition regime (region `III'), the local dimensionless strain rate $Wi$ associated with the evolution in the local minima `min1' and `min2' passes through a local maximum and then decreases until Point `P3', which corresponds to the end of region `III'. Point `P3' is characterised by a local value of the Weissenberg number, which we shall generically denote by $Wi_{EC}$, that remains approximately constant for a period of time whose duration depends on \textit{We} and $L^2$, as will be discussed below, and heralds the transition to the elasto-capillary regime in region `IV'. Towards the end of region `IV' the polymer molecules reach their maximum extensibility and the local Weissenberg number undergoes a steady increase. In this regime, beads-on-a-string structures are also formed reflecting the complex balance between capillary and elastic stresses. For the relatively large \textit{We} and small $L^2$ values used to generate Figure \ref{fig:Rmin}(b) we see that the plateau value of $Wi_{EC} \approx 1$ is only established for a narrow time interval and then the strain rate steadily increases until point 'P4', which coincides with the end of region `IV' when the finite extensibility limit is reached and the local strain rate diverges as the local radius of the thread decreases to zero.  

We note that data from the minimum radius associated with the necks established behind five different Lagrangian primary beads (labelled L $\rightarrow$ P) are shown in Figure \ref{fig:Rmin}(a,b). It is clear that the jet exhibits self-similar thinning dynamics in the initial inertio-capillary regime; this is evidenced by the overlapping curves in Figure \ref{fig:Rmin}(a) in regions `I' and `II'. If nonlinear elastic effects were not important then the linear perturbations to the jet radius would continue to grow exponentially until the radius locally approaches zero (according to Eq. (\ref{eq:LSA_Eq})) and the local strain rate $\dot{\epsilon}_{min} = -(2/R_{min})(dR_{min} / dt)$ would evolve as indicated by the curves labelled `Linear Stability' in Figures \ref{fig:Rmin}(a,b). However, as nonlinear viscoelastic effects in the fluid thread become important, the rate of thinning decreases and the local strain rate passes through a maximum labelled shortly after point 'P2'. The local thread dynamics are also self-similar in both the transition region and the elasto-capillary regime as demonstrated by the superposition of curves in regions `III' and `IV', as well as `P2' and `P3' points in Figures \ref{fig:Rmin}(a,b), respectively. It is also interesting to note that the local maximum in the strain rate associated with the minimum radius denoted `min1' (the leading ligament ahead of the forming satellite bead) is higher than the one associated with `min2', thereby highlighting the role of the momentum flux in the jet.

\begin{figure}
    \centering
    \begin{subfigure}{0.9\textwidth}
    \includegraphics[width=\textwidth]{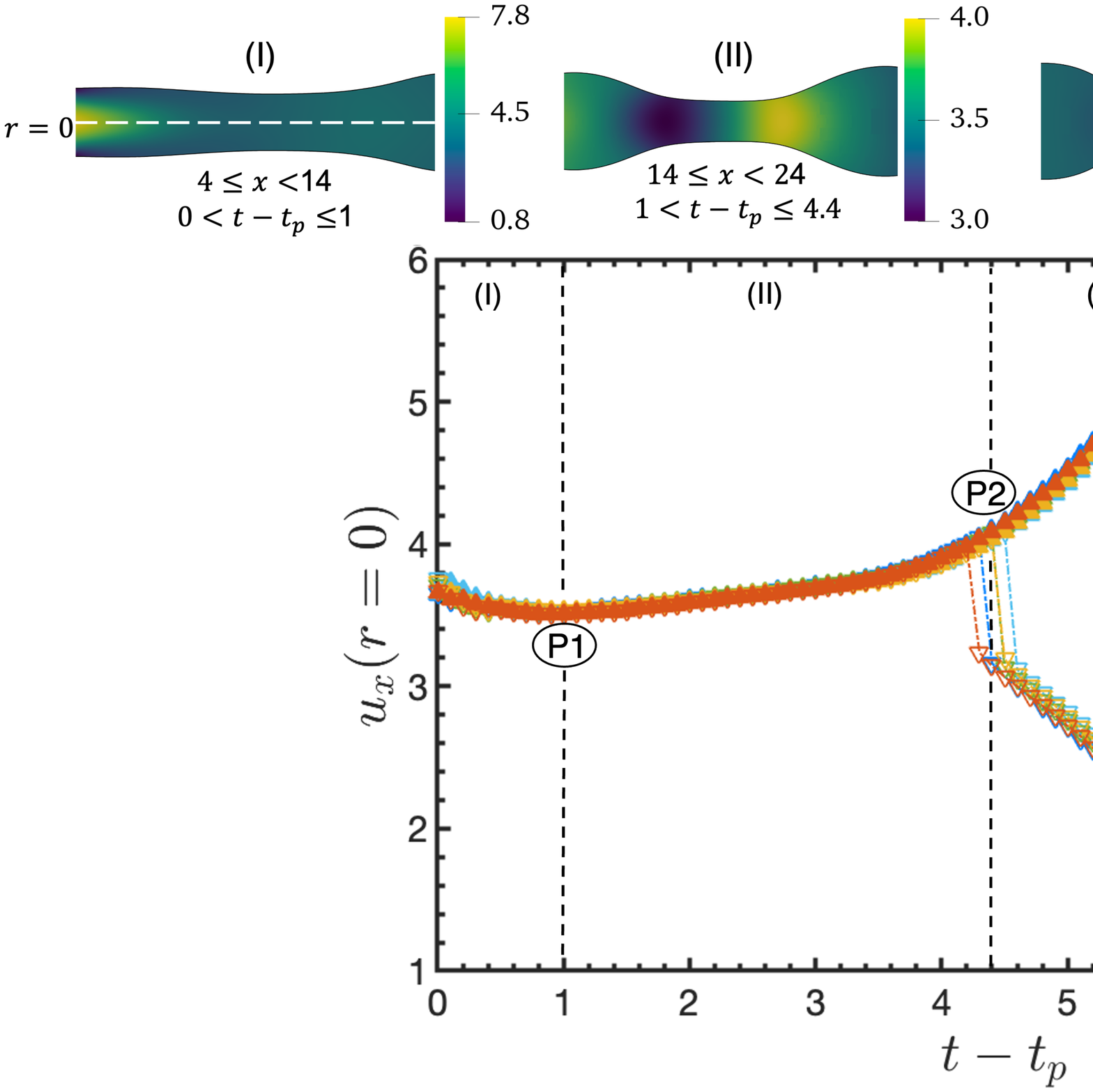}
    \caption{}
    \end{subfigure}
    \begin{subfigure}{0.9\textwidth}
    \includegraphics[width=\textwidth]{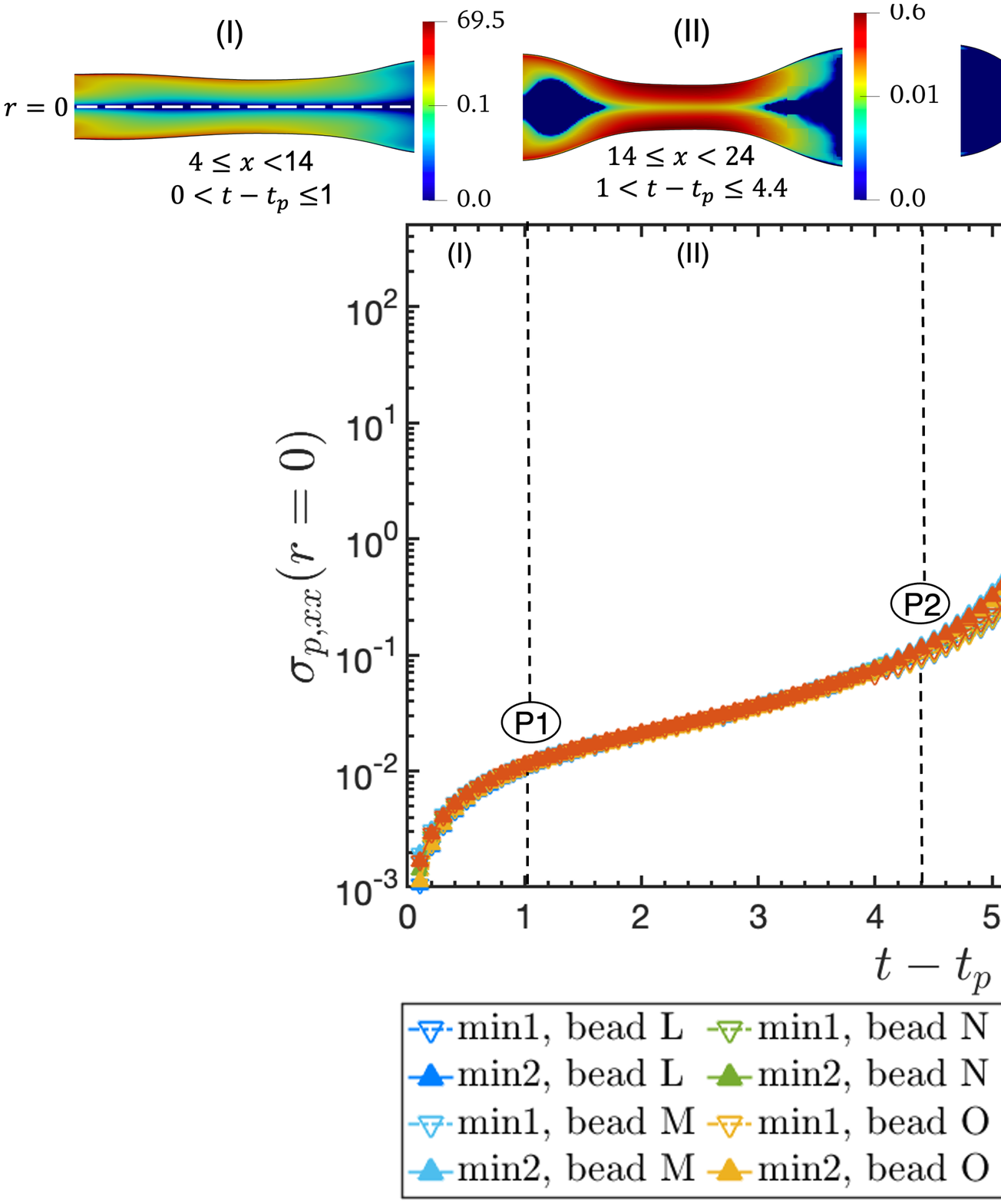}
    \caption{}
    \end{subfigure}
    \caption{Temporal variation of (a) the axial velocity component, and (b) the axial component of the elastic stress. Both are evaluated at the jet centreline using data from the local minima associated with the necks as depicted in Figure \ref{fig:Jet_LSA} considering the same simulation parameters provided in Table \ref{table_jet}. Also shown in (a) and (b) are profiles which depict the shape of the jet, coloured by contours indicating the magnitude of the axial components of the velocity and the elastic stress, respectively, during times: $0<t-t_{p} \leq 1$, $1<t-t_{p} \leq 4.4$, $4.4<t-t_{p} \leq 6.2$ and $6.2<t-t_{p} \leq 7.8$ that correspond to regions `I'-`IV' presented in Figure \ref{fig:Rmin}(a), respectively.} 
    \label{fig:Vel_Stresses}
    \end{figure}

Figures \ref{fig:Vel_Stresses}(a,b) show the dynamical evolution of the axial velocity along the centerline and the polymeric stress component at the jet centerline for two Lagrangian points corresponding to `min1' and `min2'. Also shown as inserts are snapshots of the axial velocity contours and polymeric stress fields taken at times that correspond to regions `I', `II', `III', and `IV', as defined in Figures \ref{fig:Radius_Signal} and \ref{fig:Rmin}(a).  
The centerline velocity associated with both `min1' and `min2', which correspond to the same Lagrangian element in the initial stages (regions `I'-`II'), first undergoes a decrease in region `I' as the jet exits the nozzle and the velocity profile rearranges, followed by a slow increase in region `II' associated with local perturbations growing according to the linear stability analysis. In the transition regime, region `III', the two local minima in the thread radius behind each primary bead start following different dynamics as the beads-on-a-string structure starts forming: the axial velocity associated with  point `min1' (`min2') decreases (increases) until reaching a minimum (maximum). Shortly after reaching their local extremal value, the axial velocity in Figure \ref{fig:Rmin}(a) then exhibits an approximately linear decrease (increase) during region `IV' as each part of the thinning jet approaches a constant velocity.

In contrast, the axial component of the polymeric stress shown on a semi-log scale in Figure \ref{fig:Vel_Stresses}(b) exhibits a more complex and non-uniform rise over time with high rates of increase at the exit plane of the nozzle in region `I', and in region `III' while a slower increase occurs in the region of linear instability (region `II').  
In the elasto-capillary regime (region `IV'), the exponential increase in tensile stress within the thinning filament over time is clearly seen. As expected this results in the largest values of the axial component of the polymeric stress in the thin ligament \citep{Clasen2006, eggers_herrada_snoeijer_2020, Deblais2020}. Finally beyond `P4' the finite extensibility limit is approached and the radius decreases to zero as the local strain rate and resulting stress in the thinning thread diverge.
Once again, we note that all the data obtained from the five individual Lagrangian local necked regions used to generate Figure \ref{fig:Vel_Stresses} collapse to form a master curve for the evolving axial velocity and stress components at the jet centreline highlighting the periodicity and self-similarity of the established dynamics. 

In contrast to the free viscoelastic filament undergoing thinning in the absence of a mean flow (the $We=0$ case) presented recently in \citet{Turkoz2018}, in the jetting process for a low to moderate Weber number, the momentum flux of the ejected fluid at the nozzle exit stimulates the exponential decrease of the jet radius and the development of a fore-aft asymmetric beads-on-a-string structure, as indicated by the distinct evolution of the two local minima that separate the formation of the viscoelastic ligament in Figure \ref{fig:Vel_Stresses}(a). \myRed{Below we further investigate the role of the injection flowrate and the finite-extensibility of the polymeric chains in the rate-of-thinning during the elasto-capillary regime and the resulting pinch-off dynamics.}


\subsection{\myRed{Effect of Inertia and polymer finite-extensibility}\label{sec:We_effect}}
\begin{figure}
    \begin{center} 
    \includegraphics[width=\textwidth]{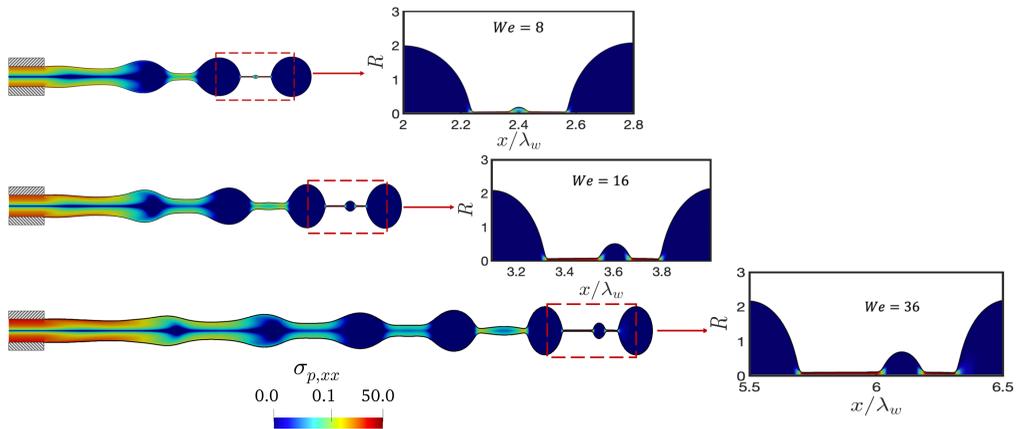}
    \caption{The effect of fluid inertia on the interfacial thinning dynamics: contour plots of the jet shape coloured by the magnitude of the axial component of the elastic stress for $We=8$, 16, and 36 at $t=53.6$, 47.8, and 29.2 respectively, with the rest of the parameters remaining unchanged from Table \ref{table_jet}. Also shown are enlarged views of the leading bead and the interconnecting ligament regions for each value of Weber number.}
    \label{fig:We_Contour_plots}
    \end{center}
    \end{figure}

In Figure \ref{fig:We_Contour_plots} we show the effect of altering the injection rate on the resulting jet dynamics by varying the Weber number; we plot snapshots of the interface shape coloured by the magnitude of the axial polymeric stress component field for $We=8$, $16$, and $36$ with the rest of the parameters remaining unchanged from those shown in Table \ref{table_jet}. The smallest Weber number studied was chosen to be larger than that associated with the so-called ``gobbling limit" (typically seen around $We \approx 2$ \citep{Clasen2009}). 
%
%
It is seen that the jet length increases with \textit{We} and the  thinning dynamics are accompanied by a concomitant rise in the number of undulations that develop into necks with longer strings separating the formed beads. According to the dispersion curves in Figure \ref{fig:nozzle_LSA}(b), the magnitude of the Weber number has a weak influence on the instability growth rates, and the perturbations are therefore advected further away from the nozzle before entering the elasto-capillary regime when the Weber number is increased. Furthermore, the size of the satellite drops along the ligaments interconnecting the primary drops also increases with \textit{We} and their position is shifted downstream towards the leading bead; the size of the primary beads, however, only appears to be weakly dependent on \textit{We}. Moreover, from the contour plots of the elastic stresses shown in Figure \ref{fig:We_Contour_plots}, it is clear that the increase in \textit{We} results in higher polymeric stresses within the nozzle and correspondingly at the exit, but these largely relax within a few jet diameters and there is only a slight increase in the stress levels attained in the thin viscoelastic threads. 

\begin{figure}
    \centering
    \includegraphics[width=0.75\textwidth,trim=0.0cm 0.0cm 1.0cm 0.2cm,clip=true]{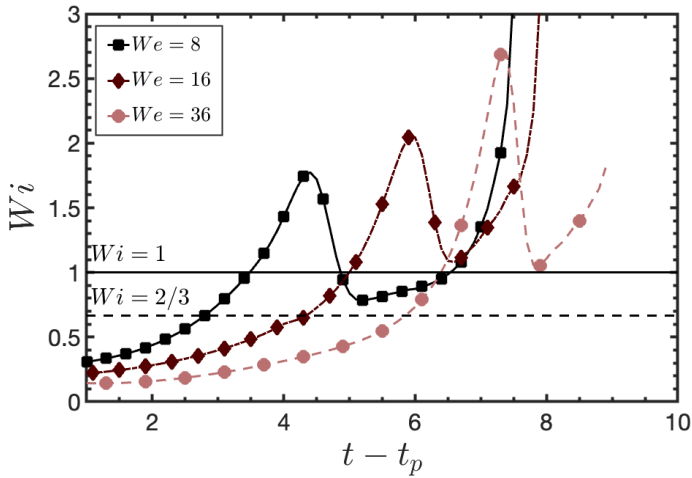}
    \caption{\myRed{The effect of increasing fluid inertia on the temporal evolution of the local dimensionless strain-rate $Wi(t) = \tau \dot{\epsilon}_{min}(t)$ in the local thinning neck of the fluid jet, and for $We=8$, 16, and 36 with the rest of the parameters remaining unchanged from Table \ref{table_jet}}.}  \label{fig:We_effect}
    \end{figure}

We also study the temporal evolution in the local dimensionless strain rate $Wi(t)=\tau \dot{\epsilon} _{min} (t)$ in Figure \ref{fig:We_effect} for Weber numbers 8, 16, and 36 with the rest of the parameters remaining unchanged from Figure \ref{fig:Vel_Stresses}. As the profiles in Figure \ref{fig:Vel_Stresses} are identical for each neck established behind a formed primary bead labelled L, M, N..., we focus only on one local neck henceforth in Figure \ref{fig:We_effect}. We also consider the flow dynamics only after dimensionless times ($t - t_p \geq 1$) during which the effect of the nozzle exit becomes less pronounced. In each case, it is clear that the evolution in the local strain rate in a fluid neck, as it evolves along the jet, shows all the features documented in Figure \ref{fig:Rmin} with a slow increase in \textit{Wi(t)} as the disturbances grow, a local maximum in the deformation rate before the necking material element enters the elasto-capillary (EC) regime (region `IV') in which the Weissenberg number approaches a locally constant value that we denote $Wi_{EC}$. However, it can be seen that decreasing the level of inertia in the jet results in the approach to a plateau value $Wi_{EC} =2/3$, in marked contrast to the $We=16$ and $36$ cases where $Wi_{EC}=1$. It is also observed that increasing the Weber number leads to larger values of the local maxima in the Weissenberg number obtained after point `P2', which coincides with the transition to the elasto-capillary regime. It is also clear that in the case of the smaller Weber number ($We=8$) the transition from the characteristic points labelled `P3' to `P4' in Figure \ref{fig:Vel_Stresses} is more gradual compared to $We=16$ and $36$. 

   %
   \begin{figure}
    \centering
    \includegraphics[width=0.77\textwidth,trim=0.01cm 0.2cm 0.6cm 0.6cm,clip=true]{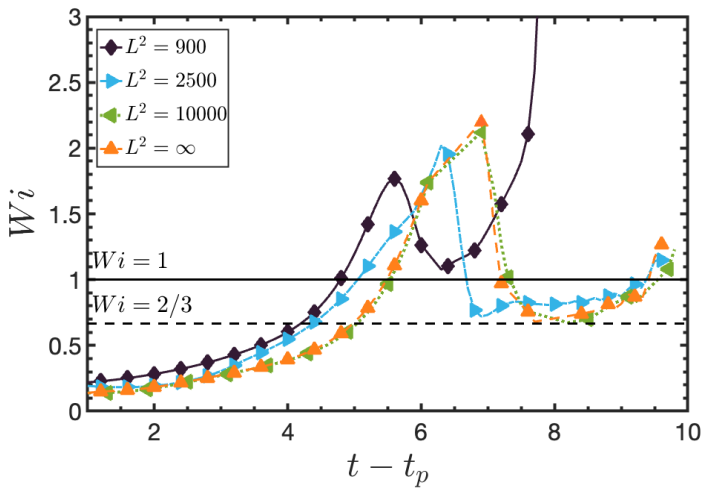}
    \caption{\myRed{The effect of increasing the extensibility parameter $L^2$ on the temporal evolution of the local dimensionless strain rate $Wi(t)= \tau \dot{\epsilon}_{min}(t)$, for $L^2=900$, 2500, 10000, and $We=16$, with the rest of the parameters remaining unchanged from Table \ref{table_jet}. Also shown for comparison the Oldroyd-B limit corresponding to $L^2 \rightarrow \infty$}.}
    \label{fig:L2_effect}
    \end{figure}

In Figure \ref{fig:L2_effect} we show how the dimensionless strain rate \myRed{in} a representative fluid neck varies with the extensibility parameter $L^2$ at a Weber number $We=16$ (the rest of the parameters remain unchanged from Table \ref{table_jet}). It is clear that an increase in the extensibility of the polymeric chains beyond $L^2 =900$ leads to the strain rate in the elasto-capillary regime converging progressively to a plateau of value \myRed{$Wi_{EC}=2/3$} for a time duration that appears to be weakly dependent on $L^2$. In contrast, for $L^2=900$, as discussed above (see Figure \ref{fig:Rmin}(b)), the limited extensibility of the chains prevents a full elasto-capillary balance from being established and there is a rapid divergence in the local strain rate from point `P3' towards `P4', with the \myRed{$Wi_{EC}=2/3$} plateau never being approached. Moreover, the local peaks in $Wi$, which coincide with the transition to the elasto-capillary balance, increase with $L^2$, reaching saturation as the Hookean dumbbell limit $L^2 \rightarrow \infty$ is approached. 

\begin{figure}
    \centering
     \includegraphics[width=1\textwidth]{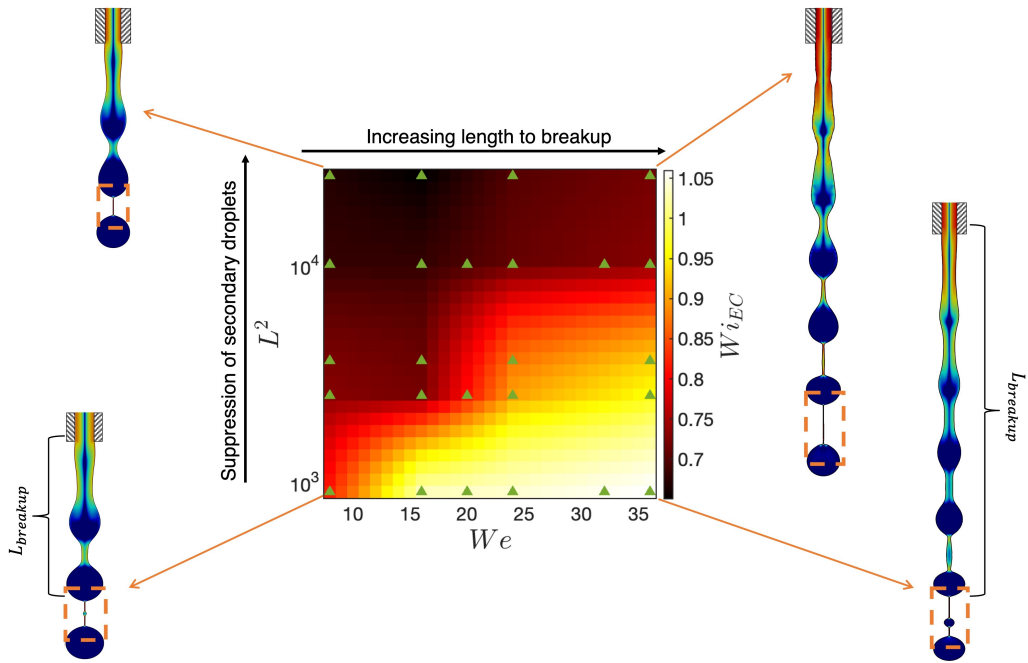}
    \caption{Flow regime map in $(We,L^2)$ parameter space coloured by the magnitude of the Weissenberg number at the start of the elasto-capillary regime, $Wi_{EC}$, which corresponds to point `P3' labelled in Figure \ref{fig:Rmin}(b). The four contour plots of the jet profile which highlight the shape of the jet for low and high values of  \textit{We} and $L^2$ are coloured by the magnitude of the axial component of the elastic stress. \myRed{The green triangles correspond to the numerical simulations that have been performed here to capture the critical regions of the $Wi_{EC}$ magnitude in the $(We,L^2)$ space.} The rest of the parameters are given in Table \ref{table_jet}. } 
    \label{fig:L2_MAP}
    \end{figure}

In Figure \ref{fig:L2_MAP} we construct a flow map in $(We,L^2)$ space in which we collect the results presented in Figures \ref{fig:We_effect} and \ref{fig:L2_effect}. The map is coloured by the magnitude of $Wi_{EC}$  established during the elasto-capillary balance. The values of $Wi_{EC}$ are computed from the strain rate in the necking filament at the onset of the elasto-capillary regime and serve to highlight whether or not the thinning dynamics are significantly accelerated beyond the value $Wi_{EC} = 2 / 3$ expected in the classic elasto-capillary balance \citep{Entov1997} depending on \textit{We} and $L^2$.

        %
Additional simulations are performed over a range of \textit{We} and $L^2$ to cover an extended region of parameter space from low to moderate jet speeds and from moderate to large polymer chain extensibilities. As indicated by the arrows in Figure \ref{fig:L2_MAP}, pronounced beads-on-a-string structures are promoted for large \textit{We}, associated with longer jet lengths with multiple beads, whilst high values of $L^2$ enable large elastic stresses to develop in the jet and lead to the formation of longer and thinner ligaments without satellite droplets attached, as well as slower thinning dynamics. When the axial momentum in the jet is small and the breakup length of the jet is constrained by a large finite extensibility, an elasto-capillary balance with $Wi_{EC} =2/3$ \myRed{(deep red colours)} is established, similar to the dynamics realised in the CaBER device \citep{Entov1997, Anna2001}. However, when the axial momentum in the jet is high, \myRed{the length to breakup is very long, and the finite-extensibility of the polymeric chains is small, the asymmetric force balance by \citet{Clasen2009} and \citet{Mathues2018} applies, resulting in a faster local stretching rate such that $Wi_{EC} \approx 1$ \myRed{(yellow colour contours)}}.


\section{Conclusions}\label{sec:Conclusions}
We have studied the thinning and breakup of an axisymmetric viscoelastic jet issuing from a nozzle using the FENE-P constitutive relation which accounts for finite polymer chain extensibility. We have used the open-source code {\it Basilisk}, which is based on a Volume-of-Fluid interface-capturing methodology and utilises adaptive mesh refinement for accurate and efficient free surface flow solutions. The free surface evolution of the jet is coupled to the upstream flow and the initial polymeric stress development inside the nozzle by employing a simplified Immersed Boundary Method. The numerical solutions of the local flow within the nozzle are in excellent agreement with analytical solutions for the cases of steady and pulsating flows \citep{WomerssleyJ.R.1955}. 

Our numerical simulations of the interfacial dynamics capture the development of beads-on-a-string structures \citep{Clasen2006} for fixed Ohnesorge and Deborah numbers over a range of Weber numbers as well as for a range of finite polymer chain extensibilities, representative of real polymer solutions. We have highlighted the development of four local regions along the jet axis. The initial growth of small amplitude perturbations is consistent with linear stability theory sufficiently close to the nozzle inlet and gives way to nonlinear dynamics further downstream where the jet evolution is first governed by the interplay of capillary, inertial, and viscous forces, and subsequently dominated by an elasto-capillary balance characterised by the formation of a distinct beads-on-a-string structure. We have also successfully captured the exponential thinning of the thin highly-stretched ligaments that form between the primary beads. This elasto-capillary thinning regime is short-lived for small polymer chain extensibilities but becomes more pronounced for high $L^2$, and the corresponding polymeric tensile stresses grow larger and larger, in the limit of infinite chain extensibility. In this limit, adaptive mesh resolution becomes essential and we are able to simulate the time-dependent evolution of the jet down to minimum feature sizes of $R_{min} \approx 0.006 R_0$ and jet lengths as large as $\ell_{jet,max} \approx 100 R_0$ (see Appendix \ref{sec:Appendix} for additional details).

Finally, we have explored in detail the local thinning dynamics of the slender ligaments that develop between the beads-on-a-string structures that evolve along the jet to resolve differences in previous reports that affect the determination of a characteristic fluid relaxation time. We construct a flow map in Weber number and chain extensibility space and calculate the variations in the local dimensionless extension rate $Wi_{EC} (We, L^2)$. This map helps us identify regions of parameter space characterised by the presence or absence of satellite drop formation, the development of very long viscoelastic jets with pronounced beads separated by thin strings, and how the thinning dynamics in the ligaments may vary from $Wi_{EC} = 2/3$ to $Wi_{EC} =1$. Understanding this systematic evolution is essential if an accurate value of the characteristic relaxation time in an unknown fluid is to be extracted from measurements of ligament thinning in a jetting rheometer or inkjetting device \citep{Morrison2010, Keshavarz2015, Mathues2018, Xu2021}.
\vspace{0.25in}

\begin{center}
{\bf Acknowledgements}
\end{center}
\vspace{0.1in}

This work is supported by the Engineering and Physical Sciences Research Council, United Kingdom, through the EPSRC PREMIERE (EP/T000414/1) Programme Grant. Support from Johnson Matthey for Konstantinos Zinelis is also gratefully acknowledged.
\\

There are no conflicts of interest to report. 


\bibliographystyle{jfm}
\bibliography{ref}

\appendix

    \begin{figure}
    \centering 
    \begin{subfigure}{0.49\textwidth}
    \includegraphics[width=\textwidth]{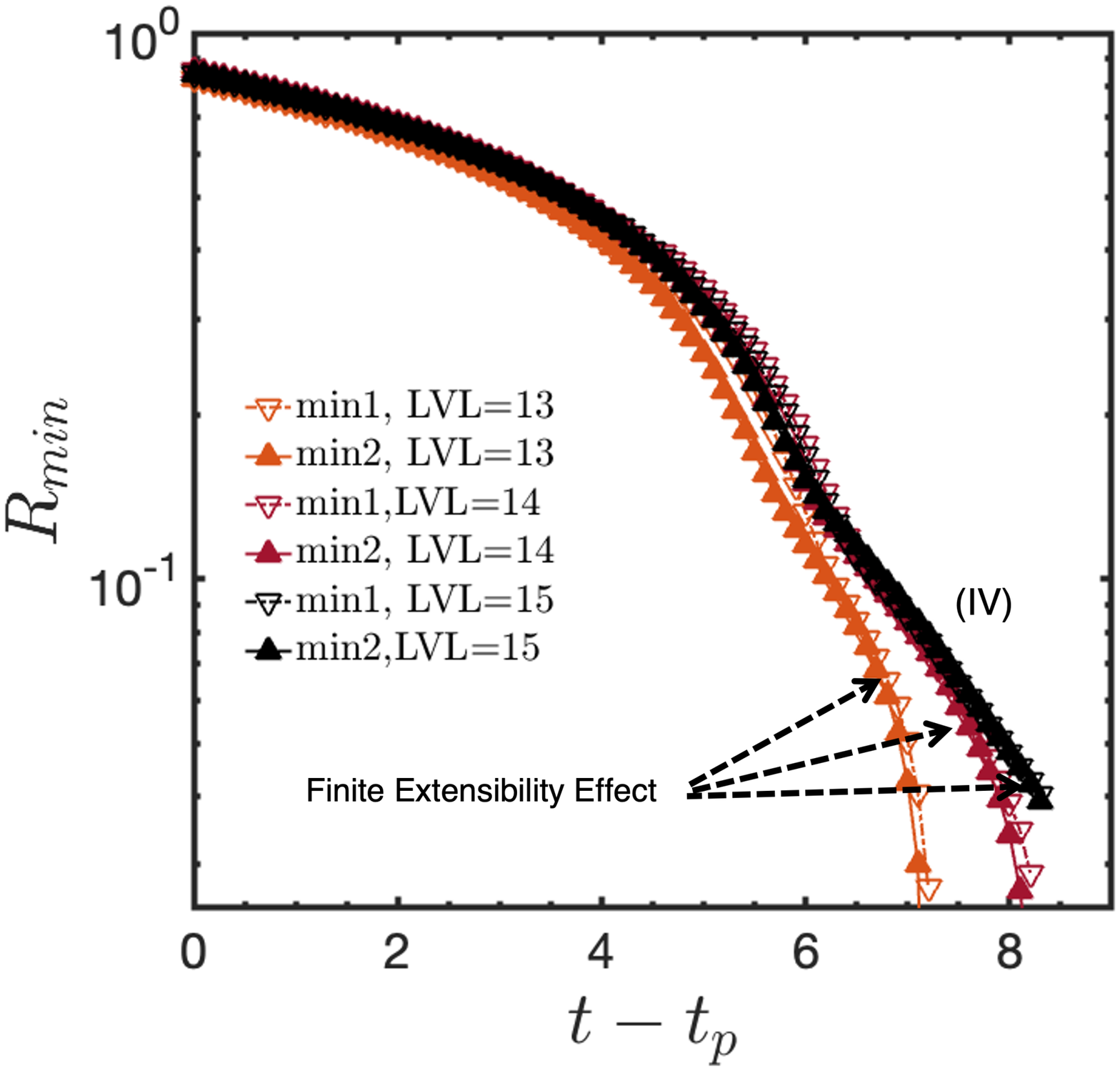}
    \caption{}
    \end{subfigure}
    \begin{subfigure}{0.49\textwidth}
    \includegraphics[width=\textwidth]{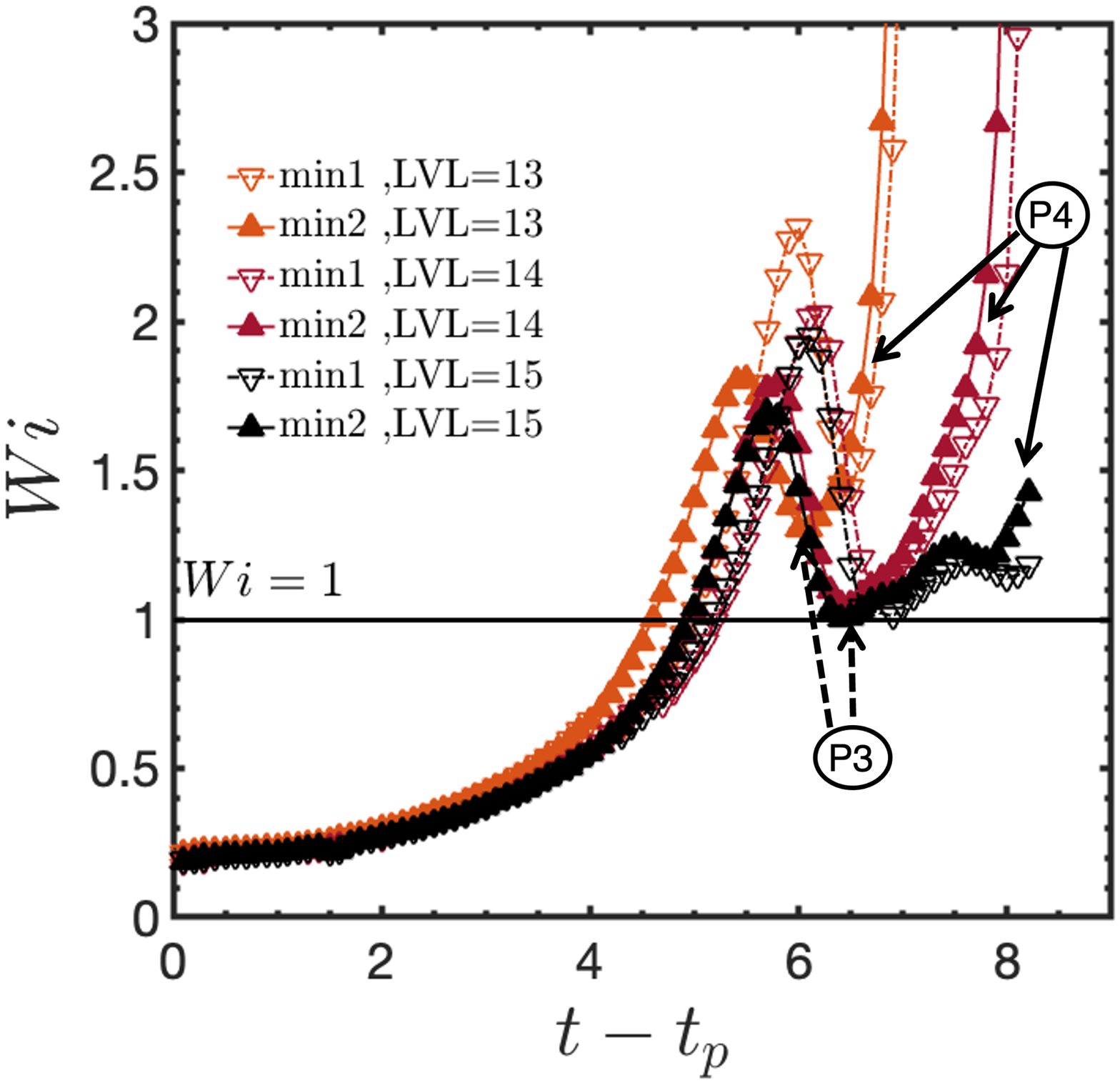}
    \caption{}
    \end{subfigure}
    \caption{Mesh convergence study showing the evolution of the minimum neck radius $R_{min}(t)$ for a representative Lagrangian element with three distinct levels-of-refinement $LVL=13,14$, and 15 for the parameters shown in Table \ref{table_jet}; results are presented through the temporal evolution of (a) the minimum radius of the jet $R_{min}$, and (b), the local dimensionless strain rate $Wi=\tau \dot{\epsilon}_{min}(t)$ to evaluate the resolution of the point `P3' where the local dimensionless strain rate $Wi(t)$ passes through a minimum.}
    \label{fig:Mesh_study}
    \end{figure}
\section{Mesh convergence study \label{sec:Appendix_B}}
It is essential to confirm that the current mesh resolution is adequate to capture all the dynamics of interest, in particular, the point `P3' labelled in Figure \ref{fig:Rmin}(b) or equivalently the value of $Wi_{EC}$ which determines the local rate of filament thinning in the elasto-capillary regime. To achieve the required level of resolution we compute and present in Figure \ref{fig:Mesh_study} the evolution in $R_{min}(t)$ and $Wi (t)$ for a lower and a higher level of refinement, $LVL=13$ and $LVL=15$, respectively. We note here that each increase of $LVL$ corresponds to a decrease in the minimum cell size by a factor of 2. For example, $LVL=14$ and $LVL=15$ correspond to $\Delta x = 0.006$ and $\Delta x = 0.003$, respectively. Figure \ref{fig:Mesh_study}(a) shows that there is a significant influence of the grid resolution on the nonlinear elasto-capillary thinning regime (region `IV'). Specifically, the local rate of thinning in the thread radius for $LVL=13$ is excessively rapid and pinch-off is approached too rapidly. The change in the slope is large as we move to a higher level of resolution ($LVL=14$). However, increasing further to $LVL=15$ does not seem to change the observed dynamics significantly, particularly regarding the local exponential decrease of the minimum radius in the elasto-capillary regime. An interesting difference between $LVL=14$ and $LVL=15$ is only observed very close to the limit of the finite extensibility of the polymer chains. A higher level of refinement delays the filament breakup and ensures a slightly longer-lasting filament as the finite extensibility effects which lead to the final breakup become significant later at $t-t_p \approx 8.4$ with $LVL=15$ compared to $t-t_p \approx 7.8$ with $LVL=14$.
Additionally, the enhanced resolution also results in a better-resolved terminal linear thinning regime where finite extensibility effects dominate. This observation can be confirmed by the evolution in the local Weissenberg number  shown in Figure \ref{fig:Mesh_study}(b). While $LVL=13$ refinement definitely leads to faster thinning dynamics, it does not allow for meaningful quantitative analysis, the higher resolution of the $LVL=14$ and $LVL=15$ simulations lead to good overlap for the two minima`min1' and `min2', in particular at point `P3'. Nonetheless, an even closer inspection shows that the `P4' point is not identical even at these two levels of resolution. Therefore, $LVL=14$ refinement is sufficient for computations of the exponential elasto-capillary regime, whereas $LVL=15$ refinement is required (but is also more computationally challenging) for analysis of the terminal finite-extensibility regime that dominates the ligament dynamics immediately before pinch-off.

\section{Linear stability analysis \label{sec:Appendix}}
Here, we provide details of the (temporal) linear stability analysis discussed in Section \ref{sec:LSA} in connection with region `II' that is identified in Figure \ref{fig:Jet_LSA}. Here we focus on the real part of the complex dispersion equation as the analysis is restricted only to examining purely temporal instabilities of the jet. Following the substitution of small amplitude perturbations in the filament radius and the linearisation of the dimensionless mass and momentum conservation equations in cylindrical coordinates, and the incorporation of the kinematic and dynamic interfacial boundary conditions, the use of normal mode analysis leads to the following dispersion relationship for an axisymmetric Oldroyd-B jet, corresponding to a dilute polymer solution with infinite chain extensibility, $L^2 \rightarrow \infty$ 
\citet{Brenn2000}: 
\begin{equation}
\begin{aligned}
&\Omega_{r}^{2} \frac{k }{2}\left[\frac{I_{0}\left(k \right)}{I_{1}\left(k \right)}+\frac{\rho_{g}}{\rho_{l}} \frac{K_{0}\left(k \right)}{K_{1}\left(k \right)}\right]+\Omega_{r} k^{2} Oh \frac{1+\beta D e \Omega_{r}}{1+ D e \Omega_{r}} \\
&\quad \times\left[2 k  \frac{I_{0}\left(k \right)}{I_{1}\left(k \right)}\left(1+ k^{2} \frac{O h}{\Omega_{r}} \frac{1+\beta De \Omega_{r}}{1+D e \Omega_{r}}\right)\right. \\
&\left.-1-2 l  \frac{I_{0}\left(l \right)}{I_{1}\left(l \right)} k^{2} \frac{O h}{\Omega_{r}} \frac{1+\beta D e \Omega_{r}}{1+D e \Omega_{r}}\right] \\
&\quad=\frac{k^{2}}{2}\left(1- k^{2}\right)+C \frac{\rho_{g}}{\rho_{l}} k^{3} We \frac{K_{0}\left(k \right)}{K_{1}\left(k \right)}.
\end{aligned}
\label{eq:LSA_Appendix}
\end{equation}
Here, $\Omega_{r}>0$ ($\Omega_r<0$) indicates instability (stability), $I_n$ and $K_n$ are the modified Bessel functions, $C$ is an empirical correction factor to express the aerodynamic effects on the jet (here we choose $C = 0.175$ \citep{Brenn2000,Keshavarz2015}), and $l$ is a dimensionless modified wavenumber given by $l^{2} = k^{2}+\left(1+De\left(\Omega +i k {U}_{0} \right) \right) / \left( Oh (1+\beta De \left(\Omega+i k U_{0} \right)) \right)$. The intrinsic Deborah number is $De = \tau/t_{R}$, the Ohnesorge number is $Oh= \eta_0 / \sqrt{\rho R_0 \gamma}$ and the solvent viscosity ratio is $\beta = \eta_s / \eta_0$, with the Rayleigh time $t_{R}= \sqrt{\rho R_{0}^3 / \gamma}$ and the initial radius of the jet $R_{0}$ being the characteristic time and length scales, respectively. Eq. (\ref{eq:LSA_Appendix}) can be solved numerically with a simple MATLAB-solver at a specific value of $De$, $Oh$ and \textit{We} for a range of wavenumbers with the dimensionless growth rate $\Omega_r$ being the unknown. Typical results are presented in Figure \ref{fig:Jet_LSA}(b): In the inviscid limit the maximum growth rate $\Omega_r \approx 0.34$ is at $k \approx 0.693$. The first effect of viscosity is to stabilise the jet partially and the most unstable growth rate reduces to $\Omega_r \approx 0.23$ at $k \approx 0.6$. In contrast, linear viscoelastic effects are observed to render the jet slightly more unstable than the corresponding Newtonian viscous jet with the most unstable growth rate increasing to $\Omega_r \approx 0.24$ at $k \approx 0.6$.



\end{document}